\documentclass[showpacs,prb,twocolumn]{revtex4}%,preprint,draft
\usepackage{amsmath}
\usepackage{amssymb}
\usepackage{graphicx}
\usepackage{dcolumn}
\usepackage{bm}

\def\a{^{\ast}}
\def\q{\quad}

\def\c{ \cdot }

\def\p{ \partial }

\def\a{\alpha}%
\def\Om{\Omega}%
\def\G{\Gamma}%
\def\mD{\mathcal{D}}%
\def\p{ \partial }%

\def\equ{Eq.~\eqref}

\def\boldrm#1{{\bf #1}}
\def\boldit#1{\mbox{\boldmath$#1$}}
\makeatletter
 \def\eqalign#1{\null\vcenter{\def\\{\cr}\openup\jot\m@th
 \ialign{\strut$\displaystyle{##}$\hfil&$\displaystyle{{}##}$\hfil
  \crcr#1\crcr}}\,}

\begin{document}

\title{Excitation of surface plasmon-polaritons in metal films with
double\\ periodic modulation: anomalous optical effects}

\author{A. V. Kats$^1$}
 \email{ak_04@rambler.ru}
\author{M. L. Nesterov$^{1,2}$}
 \email{nesterovml@web.de}
\author{A. Yu. Nikitin$^{1,3}$}
\email{alexeynik@rambler.ru}

\affiliation{$^1$ Theoretical Physics Department, A.Ya. Usikov Institute for Radiophysics and Electronics, Ukrainian
Academy of Sciences, 12 Acad. Proskura Str., 61085 Kharkov, Ukraine\\
$^2$Photonics Research Group, Aston University, Birmingham B4 7ET, United Kingdom\\
$^3$ Departamento de F\'{i}sica de la Materia Condensada-ICMA, Universidad de Zaragoza, E-50009 Zaragoza, Spain}

\date{\today}

\begin{abstract}
We perform a thorough theoretical analysis of resonance effects when an arbitrarily polarized plane monochromatic wave
is incident onto a double periodically modulated metal film sandwiched by two different transparent media. The proposed
theory offers a generalization of the theory that had been build in our recent papers for the simplest case of
one-dimensional structures to two-dimensional ones. A special emphasis is placed on the films with the modulation
caused by cylindrical inclusions, hence, the results obtained are  applicable to the films used in the experiments. We
discuss a spectral composition of modulated films and highlight the principal role of ``resonance'' and ``coupling''
modulation harmonics. All the originating multiple resonances are examined in detail. The transformation coefficients
corresponding to different diffraction orders are investigated in the vicinity of each resonance. We make a comparison
between our theory and recent experiments concerning enhanced light transmittance (ELT) and show the ways of increasing
the efficiency of these phenomena. In the appendix we demonstrate a close analogy between ELT effect and peculiarities
of a forced motion of two coupled classical oscillators.
\end{abstract}


\maketitle

\section{\label{sec:Intro}Introduction}

It is exactly tangible advances in structuring metals on nanoscales that give rise to a grate amount of experimental
and theoretical works in the field of plasmonics, which examine resonance optic effects caused by surface plasmon
polaritons (SPPs) excitation in structured conducting or semiconducting
media\cite{Book_Raether,Book_Agranovich,Zayats_Maradudin_2005}. These quasi two-dimensional (2D) electrodynamic objects
have been thoroughly studied in the solid state physics, physics of surfaces and diffraction optics over the late few
decades\cite{Ebbesen_Nature_03}. An SPPs are electromagnetic surface waves, coupled to the collective electron
excitation, they attract a great deal of attention due to their unique possibility of light localization and essential
enhancement of the electric field near the surface.

A considerable interest in SPPs stems from the latest experiments on ELT phenomena. Since 1998, after observation by
Ebbesen et.al.\cite{Ebbesen98_Nature,Ebbesen_PRB_98} the violation of Bethe's approach\cite{Bethe} to the diffraction
by subwavelength periodic hole arrays in real metal films, this has been the subject of many studies. Until the recent
moment one of the most generally recognized explanations of the ELT through subwavelength periodic hole arrays is the
excitation of SPPs. Most of the authors (especially theoreticians), who hold on to the SPP conception of ELT, assume
that the field enhancement results from ``interface'' SPPs, being either single-boundary localized (which is a common
case for a nonsymmetric dielectric surrounding of the film) or double-boundary localized for the symmetric surrounding.
A periodic hole array plays a role of a coupler between the SPPs existing at the boundaries of the conducting film and
the incident light. In this context the crucial point is the surface periodicity. The periodicity caused by other
factors, such as corrugation or periodic modulation of the medium electromagnetic properties, etc., also result in the
light-matter interaction resonance features, in particular, the ELT effect. Note that comparing the ELT transmittance
peaks positions to those caused by SPP excitation in different diffraction orders undoubtedly points out a significant
SPP role, see numerous experiments, e.g.
Refs.~\onlinecite{Ebbesen98_Nature,Ebbesen_PRB_98,Ebbesen2001_Opt,Ebbesen_theory_PRL01,
StrongCoupling_PRB05,shadows_APL_02,Fluorescence_OptLett_03,Barnes_PRB04,hole_depth_APL02}.

It should be noted that the observed ELT effects  are strongly dependent upon the film surrounding. As the film having
the subwavelength hole array is surrounded by the dielectrics with the same dielectric constants (for instance, a
free-standing film) the excited SPPs are double-boundary-localized, that is, the field is enhanced at both faces of the
film, see experiments\cite{Ebbesen2001_Opt,Ebbesen_theory_PRL01}. In this geometry the ELT in zeroth diffraction order
is much more pronounced as compared to the nonsymmetrical film surrounding, when the excited SPPs are
single-boundary-localized, that is the field can be enhanced at a single face of the film. Generaly, this corresponds
to the films deposited onto the quartz substrate. Both of these cases have been so far studied by using numerical
methods\cite{Ebbesen_theory_PRL01,Ebbesen2001_Opt,WoodHoles_PRB03,Popov_PRB00,shadows_APL_02,
Book_Petit03,Zayats_PRL01,plasmonic_metamaterials_JOPA05}. It was also shown numerically and experimentally that the
ELT occurs for periodical conducting structures without holes. Basically this is quite evident, since the type of
periodicity does not play a crucial role for the excitation of the interface SPPs. These were chiefly the structures
with relief corrugations of the film faces both for symmetrical\cite{Bloch_mode_OSA04} and
nonsymetrical\cite{Grating_couplers_99,Arutsky_OptLett_00,Nonzeroth_Barnes_APL_01} dielectric surrounding and the
structures with periodically-located dielectric pillars\cite{Popov2003_OptExpr}.

Several authors have developed an analytical approach which qualitatively describes the ELT. They have studied the
diffraction by the film with one-dimensional (1D) periodically modulated dielectric
permittivity\cite{LightTunelling_PhysRev,Dykhne_PRB_03,AnalyticalTheory_PhysRev,
Genchev_JETP04,Bliokh05,Franches_OptExpr05}. In these works a study has been made in the simplest case of a strictly
normal incidence onto the symmetrically surrounded film with harmonic modulation of the film permittivity. The
exception are the papers \cite{AnalyticalTheory_PhysRev,Franches_OptExpr05}, where the theory is generalized to the
nonsymmetric surrounding. The authors of these works have described the zeroth-order transmittance dependence upon the
parameters in case of SPP excitation  in the first diffraction order. In contrast to them, in a certain sense, we have
provided a more general analytical insight\cite{Ours_JETPHL_04,film_PhysRev,KNN_PhysRev}, by solving the problem of
vector diffraction by the film with nonsymmetric and symmetric dielectric surrounding for arbitrary modulation Fourier
spectra of the permittivity of the metal at an arbitrary incident angle and at an arbitrary polarization of the
incident light. The advantage of our analytical treatment is that we have described not only the first-order
resonances, but have given a classification of the resonances corresponding to single or double-boundary-localized SPPs
excitation in single or multiple diffraction orders. Also, we have examined a nonzeroth-order ELT that was observed in
the experiments Refs.~\onlinecite{Nonzeroth_Barnes_APL_01,Korea_03}. Mention also paper \cite{LeePark_Nanoslits_PRB05},
where the alternative analytical approach was suggested to describe the light transmittance through metallic nanoslit
structures.

Yet another important class of plasmonic structures possessing interesting diffraction resonance phenomena are metallic
nanoparticle arrays\cite{GoldNanoparticlePlasmon_NANOL05,Nano_Chem_01,NaomiHalas_MRS05,Barnes_PRB04}. As known, they
can support the so-called \emph{localized SPP resonances} which are strongly dependent upon the shape of the individual
particles\cite{NaomiHalas_MRS05}. It is believed that the \emph{shape resonances} similar to the localized ones, may
affect the ELT in subwavelength hole arrays. The influence of the hole or/and nanoparticle shape was studied
experimentally in\cite{LocalizedSPP_JOPA05,hole_shape_PRL04,shape_localized_PRB05}. The study of the conducting films
containing periodic lattices of dielectric nanoparticles and voids have been made in Refs.~\onlinecite{Gang_Sun_2006,
Kelf_2006,local_SPP_in_voids}. However, we are not aware of the study on the experimental or theoretical optical
properties of metallic nanoparticle arrays immersed into a conducting film. Similarly to a nanohole array, such a
nanoparticle array should display ELT with the wavelength spectra depending strongly upon the nanoparticle shape.

In the present paper we go on investigating the resonance optical effects by generalizing the earlier developed
analytical treatment\cite{film_PhysRev,Ours_JETPHL_04} of the conducting films with 2D modulation. We consider the
vector diffraction problem for periodically located metallic inclusions in the metal film with an arbitrary dielectric
surrounding. Since the inclusions are presumed to be entirely imbedded into the film (the faces of the film being
flat), the resonances of our system correspond to the excitation of purely interface SPPs, the localized-SPP resonances
do not exist. Along with the fact that such structures are of interest by themselves, they may be considered as a model
of subwavelength hole arrays; and the approach developed is appropriate for other periodical structures (say, for
corrugated films). One more substantial aspect of our approach is that we can easily describe the polarization of light
transmitted or/and reflected by the 2D periodical structure. The polarization properties of subwavelength hole arrays
is also a matter of interest, and they have been intensively studied in recent years\cite{StrongPolariz_PRL04,
PolarizControl_OptLett04,PolarizTomography_OptLett05, Opt.Depolariz_PRB05,terahertz_Appl.Opt._04,KNN_PhysRev}.

The paper is arranged as follows. Following the Introduction, in Section~\ref{sec:general}, we describe a general
approach to the problem of resonance light diffraction by a 2D periodically-modulated conducting film for the conical
mount case and for an arbitrary polarization of the incident light. We accentuate the fact that the form of the
periodically-located inclusions have an impact upon the Fourier spectra of the periodical structure and this, in its
turn, influences the excited SPPs considerably. In Section~\ref{sec:nonsymmetric} we examine the excitation of
single-boundary-localized SPPs in the nonsymmetrically-sandwiched film, considering both the entire transmittance and
reflectance spectra and a more detailed analysis of different resonances in their close vicinity. We give an
explanation of the recently observed polarization dependence upon the hole shape\cite{hole_shape_PRL04} from the view
point of Fourier spectra of the periodical structure. Besides, we compare our calculations with other recent
experiments. Section~\ref{sec:symmetric} is devoted to the resonance effects caused by the excitation of
double-boundary-localized SPPs in the symmetrically-sandwiched film. We study in detail the fine structure of the
two-humped resonance maxima (minima) of the transmittance (reflectance), stressing the role of long-range and
short-range SPPs. As far as we know, the fine structure of two-humped resonances corresponding to the symmetrically
surrounded film has not been studied thoroughly in experiments. In the Appendix below we draw an analogy between the
ELT and the forced oscillations of a well-known classical system of two weakly coupled linear damping mechanical
oscillators.

The problem under consideration is of profound interest not only from the view point of a pure physics, but can find
numerous applications in optical subwavelength devices design\cite{Ebbesen_Nature_03,Plasmon_devices_05}. For instance,
recently, SPP in sandwiched structures found itself as surface plasmon resonance optical
sensors,\cite{Jiri_Homola_1999, Jiri_Homola_2003, Paul_V_Lambeck_2006} and some of them are already commercially
available. These sensors are based on the interaction between waveguide and SPP  modes. The basic feature of the
sensors is measuring the refractive index variation in biological and chemical real-time high-definition sensing. It is
important to mention that ELT phenomena are studied in the microwave (THz)  region of electromagnetic waves
\cite{ELT_microwave_JOPA05,THz_SPP_diploma03,terahertz_Appl.Opt._04} which attracts a great deal of attention.

\section{\label{sec:general}Problem statement and main equations}

\subsection{\label{subsec:equations}Analytical approach}
\begin{figure}[!htb]
  \includegraphics[width=8cm]{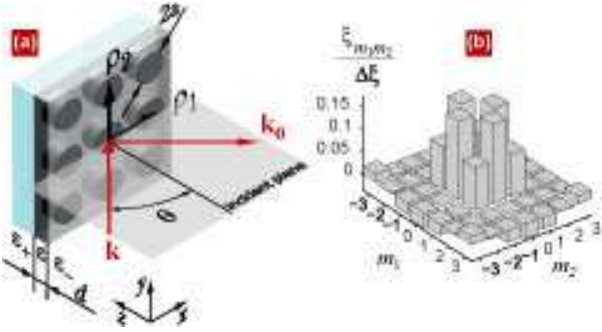}\\
  \caption{\label{fig:geometry} (Color online)(a)
Geometry of the problem. Diffraction by the nonsymmetrically-sandwiched 2D modulated
metal film (periodical modulation is due to the cylindrical inclusions). (b) Fourier
spectrum of a periodical array with $\rho_1=\rho_2=\rho$, $a/\rho=1/3$ shown in (a), see
Eq.~\eqref{ns1}.}
\end{figure}

Consider an arbitrary polarized plane monochromatic wave with  wave vector $\boldrm{k}$ incident onto a surface of a
double periodically modulated metal film surrounded by dielectric media with permittivities $\varepsilon_{\tau}$, $\tau
= \pm$, from the medium corresponding to $\tau = -$. Imply that the periodicity is caused by modulation of the
dielectric permittivity of a conductor, $\varepsilon=\varepsilon(\mathbf{r}_{t})$, $\mathbf{r}_{t} =(x,y)$, so that
$\varepsilon\left(\mathbf{r}_{t}\right)=\varepsilon \left(\mathbf{r}_{t}+m_1\boldit{\rho}_1+m_2\boldit{\rho}_2\right)$,
$\boldit{\rho}_{1,2}$ are elementary translation vectors, see Fig.~\ref{fig:geometry}.
 In what follows one has to deal with the Fourier expansion of the function
$\xi(\mathbf{r}_{t})=\sqrt{\bar{\varepsilon}}/\varepsilon(\mathbf{r}_{t})$,  where $\overline{\varepsilon}$ is the mean
value of the dielectric permittivity of the metal, $\overline{\varepsilon}=\langle\varepsilon(\boldrm{r})\rangle$. The
expansion of this function over $[\overline{\varepsilon}-\varepsilon(\mathbf{r}_{t})]/\overline{\varepsilon}$ coincides
in the zeroth order with the expansion of the surface impedance, $1/\sqrt{\varepsilon(\mathbf{r}_{t})}$, over the same
parameter. Therefore, let us recall, for brevity, the function $\xi(\mathbf{r}_{t})$ surface impedance. We write its
Fourier representation as follows
\begin{equation}\label{0}
\xi(\boldrm{r}_{t}) = \xi_{\mathcal{O}} +
\sum_{\mathcal{M}}\tilde{\xi}_{\mathcal{M}}\exp[i(m_1\boldrm{g}_1+m_2\boldrm{g}_2)\boldrm{r}_{t}] \, ,
\end{equation}
where $\tilde{\xi}_{\mathcal{O}}=0$, $\mathcal{M}$ is a vector index, or a multiindex, $\mathcal{M}=(m_1,m_2)$
(integers $m_{1}$, $m_{2}$ indicate a number of elementary translations along vectors of reciprocal grating,
$\mathbf{g}_1$, $\mathbf{g}_2$); the zero multiindex is $\mathcal{O}\equiv(0,0)$. The electric fields in the dielectric
media are written in the form of Rayleigh expansion [the time dependence $\exp \left(-i\omega t\right)$ is omitted
everywhere]:
\begin{eqnarray}\label{1}
&&\boldrm{E}^{\tau}(\boldrm{r}) = \delta_{\tau,-}\boldrm{E} \exp(i\boldrm{k}\boldrm{r})
+\nonumber\\
&&\sum_{\mathcal{M}}\boldrm{E}^{\tau}_{\mathcal{M}}
\exp\left[i\boldrm{k}_{\mathcal{M}t}\boldrm{r}_{t}+ i\tau
k_{\tau|\mathcal{M}z}(z-\delta_{\tau,+}d)\right] ,
\end{eqnarray}
for $z\geq d$ $(z\leq0)$ if $\tau=+(-)$. Here $d$ is the film thickness,
$\mathbf{E}$ denotes the electric field amplitude of the incident wave,
$\boldrm{r}=(x,y,z)$, tangential, $\mathbf{k}_{t}$,
$\boldrm{k}_{\mathcal{M}t}$, and normal, $k_{z}$, $k_{\tau|\mathcal{M}z}$,
components of the wave vectors of spatial field harmonics,
$\boldrm{k}=\boldrm{k}_{t} + \boldrm{e}_{z}k_{z}$,
$\boldrm{k}_{\mathcal{M}}^{\tau}=\boldrm{k}_{\mathcal{M}t}+\boldrm{e}_{z}k_{\tau|\mathcal{M}z}$,
are
\begin{eqnarray}\label{2}
    \nonumber
k_{z} = k_{-}  \cos\theta, \; \boldrm{k}_{t} = k_{-} (\sin\theta, 0,0) , \; \\
\nonumber
 \boldrm{k}_{\mathcal{M}t} = \boldrm{k}_{t} + m_1\mathbf{g}_1+m_2\mathbf{g}_2,
\; \\
k_{\tau|\mathcal{M}z}=\sqrt{k^{2}_{\tau}-\boldrm{k}_{\mathcal{M}t}^2} \, , \;
k_{\tau} = \sqrt{\varepsilon_{\tau}} k , \q k = \omega/c ,
\end{eqnarray}
where $\theta$ is an angle of incidence, $\Re e (k_{\tau|\mathcal{M}z}), \Im m (k_{\tau|\mathcal{M}z}) \geq0$.
Analogously to \cite{film_PhysRev,Ours_JETPHL_04,KNN_PhysRev} we will take into account the modulation in the boundary
conditions only, so that within the conducting film we seek the solution in the form
\begin{eqnarray}\label{3}
&& \overline{\mathbf{E}}(\boldrm{r}) =
 \sum_{\mathcal{M}, \tau}\overline{\boldrm{E}}^{\tau}_{\mathcal{M}}
 \exp(\tau\tilde{k}z + i\boldrm{k}_{\mathcal{M}t}\boldrm{r}_{t}), \nonumber\\
 && \tilde{k} =
 k\sqrt{-\overline{\varepsilon}},
  \quad 0 \le z \le d .
\end{eqnarray}
Introducing polarization unit vectors:
 \begin{eqnarray}\label{11}
&\boldrm{e}_{\mathcal{M}}^{\tau|+} =
\dfrac{\boldrm{e}_{z}\times\boldrm{k}_{\mathcal{M}t}}{k_{\mathcal{M}t}} \, , \q
 \boldrm{e}_{\mathcal{M}}^{\tau|-} =
\dfrac{\boldrm{e}_{\mathcal{M}}^{\tau|+}\times \boldrm{k}^{\tau}_{\mathcal{M}}}{k_{\tau}}
\, ,\nonumber\\
&\boldrm{e}^{-|+} =\dfrac{\boldrm{e}_{z}\times\boldrm{k}_{t}}{k_{t}} \, , \q
 \boldrm{e}^{-|-} =
 \dfrac{\boldrm{e}^{-|+}\times\boldrm{k}}{k_{-}} \, ,
\end{eqnarray}
where $\boldrm{e}_{\mathcal{M}}^{\tau|\sigma}$ correspond to $TM$ ($TE$) or $p$ ($s$) polarization for $\sigma=-(+)$ in
$\mathcal{M}$th diffraction order in the dielectric media $\tau$, and $\boldrm{e}^{-|\sigma}$ are polarization basis
vectors for the incident wave.  With respect to the polarization vectors, the electric and magnetic fields in the
dielectric media are
\begin{equation}\label{13}
\left[%
\begin{array}{cc}
  \boldrm{E}_{\mathcal{M}}^{\tau} \\
  \boldrm{H}_{\mathcal{M}}^{\tau} \\
\end{array}%
\right] =  \sum_{\sigma}
\left[%
\begin{array}{cc}
  E_{\mathcal{M}}^{\tau|\sigma} \\
  H_{\mathcal{M}}^{\tau|\overline{\sigma}}\\
\end{array}%
\right] \boldrm{e}_{\mathcal{M}}^{\tau|\sigma}, \;\;
\left[%
\begin{array}{cc}
  \boldrm{E} \\
  \boldrm{H} \\
\end{array}%
\right]
 =  \sum_{\sigma}
 \left[%
\begin{array}{cc}
   E^{\sigma} \\
   H^{\overline{\sigma}} \\
\end{array}%
\right]\boldrm{e}^{-|\sigma}\! \! ,
\end{equation}
where $\overline{\sigma}\equiv-\sigma$; and analogously inside the film. Let us
introduce the polarization transformation coefficients (TCs),
$T^{\tau|\sigma\sigma'}_{\mathcal{M}}$, as
\begin{equation}\label{37}
  \begin{split}
E^{\tau|\sigma}_{\mathcal{M}}=
\sum_{\sigma'}T^{\tau|\sigma\sigma'}_{\mathcal{M}}E^{\sigma'}.
\end{split}
\end{equation}
Then, excluding from the Maxwell equations and the boundary conditions the
internal fields, we arrive at the following infinite linear system for the TCs
corresponding to the outer fields
\begin{equation}\label{38}
  \begin{split}
&\sum_{\mathcal{M}',\tau',\sigma''}
D_{\mathcal{M}\mathcal{M}'}^{\tau\tau'|\sigma\sigma''}
T^{\tau'|\sigma''\sigma'}_{\mathcal{M}'}= V_{\mathcal{M}}^{\tau|\sigma\sigma'}.
\end{split}
\end{equation}
The matrix $D_{\mathcal{M}\mathcal{M}'}^{\tau\tau'|\sigma\sigma''}$ and the
right-hand side vectors $V_{\mathcal{M}}^{\tau|\sigma\sigma'}$ are linear with
respect to the modulation,
\begin{eqnarray}\label{43}
&&D_{\mathcal{M}\mathcal{M}'}^{\tau\tau'|\sigma\sigma'}=\delta_{\mathcal{M}\mathcal{M}'}
\delta_{\sigma\sigma'}b_{\mathcal{M}}^{\tau\tau'|\sigma}+
d_{\mathcal{M}\mathcal{M}'}^{\tau\tau'|\sigma\sigma'}, \nonumber\\
&&V_{\mathcal{M}}^{\tau|\sigma\sigma'}=\delta_{\mathcal{M},\mathcal{O}}
\delta_{\sigma\sigma'}V_{\mathcal{O}}^{\tau|\sigma}+v_{\mathcal{M}}^{\tau|\sigma\sigma'}.
\end{eqnarray}
The modulation-independent terms are diagonal both with respect to the
diffraction order and polarization. Explicitly,
\begin{widetext}
\begin{eqnarray}\label{44}
  &&b_{\mathcal{M}}^{\tau\tau'|\sigma}=\tau'\sigma\left[\delta_{\tau\tau'}
\varepsilon_{\tau'}^{-\frac{1+\sigma}{4}}
\beta_{\tau'|\mathcal{M}}^{\frac{1-\sigma}{2}}\mathrm{th}\Phi
+\left(\overline{\sigma}\right)^{\frac{1-\tau'}{2}}\tau^{\frac{\sigma+1}{2}} \xi_{\mathcal{O}}
(\beta_{\tau'|\mathcal{M}}\sqrt{\varepsilon_{\tau'}})^{\frac{1+\sigma}{2}} \right]
\left(\mathrm{cosh}\Phi\right)^{\frac{1+\tau\tau'}{2}},\nonumber\\
 &&V_{\mathcal{O}}^{\tau|\sigma}=
\left[\delta_{\tau,-}\varepsilon_{-}^{-\frac{1+\sigma}{4}}
\beta_{-|\mathcal{O}}^{\frac{1-\sigma}{2}}\mathrm{th}\Phi + \sigma\tau^{\frac{\sigma+1}{2}}
 \xi_{\mathcal{O}}
(\beta_{-|\mathcal{O}}\sqrt{\varepsilon_{-}})^{\frac{1+\sigma}{2}}\right]
\left(\mathrm{cosh}\Phi\right)^{\frac{1-\tau}{2}} .
\end{eqnarray}
\end{widetext}
Here
\begin{equation}\label{23.1}
  \begin{split}
\Phi\equiv\tilde{k}d,
\end{split}
\end{equation}
the real part of $\Phi$ is the film thickness in the skin-depths, $\beta_{\tau|\mathcal{M}}$ is the dimensionless $z$
component of the $\mathcal{M}$th field harmonic wave vector, which is related with the coresponding tangential
component $\boldit{\alpha}_{\mathcal{M}}$ as
\begin{equation}\label{35}
  \begin{split}
\beta_{\tau|\mathcal{M}}=\frac{k_{\tau|\mathcal{M}z}}{k\varepsilon_{\tau}}=
\frac{\sqrt{\varepsilon_{\tau}-\alpha_{\mathcal{M}}^{2}}}{\varepsilon_{\tau}} \, , \q
\boldit{\alpha}_{\mathcal{M}}=\boldrm{k}_{\mathcal{M}t}/k.
\end{split}
\end{equation}
The nondiagonal terms are
\begin{eqnarray}\label{45}
d_{\mathcal{M}\mathcal{M}'}^{\tau\tau'|\sigma\sigma'}= \tau^{\frac{\sigma+1}{2}}
\sigma^{\frac{1-\sigma'}{2}}\left(\sigma'\right)^{\frac{1-\tau'}{2}}
S_{\mathcal{M}\mathcal{M}'}^{\sigma\cdot\sigma'}
 \tilde{\xi}_{\mathcal{M}-\mathcal{M}'}\nonumber\\
\times(\beta_{\tau'|\mathcal{M}'}\sqrt{\varepsilon_{\tau'}})^{\frac{1+\sigma'}{2}}
\left(\mathrm{cosh}\Phi\right)^{\frac{1+\tau\tau'}{2}},\nonumber\\
v_{\mathcal{M}}^{\tau|\sigma\sigma'}=
\sigma'd_{\mathcal{M}\mathcal{O}}^{\tau-|\sigma\sigma'}.
\end{eqnarray}
$S_{\mathcal{M}\mathcal{M}'}^{+}$ ($S_{\mathcal{M}\mathcal{M}'}^{-}$) are sines
(cosines) of the angle
$\psi_{\mathcal{M}\mathcal{M}'}=\left(\widehat{\boldit{\alpha}_{\mathcal{M}},
\boldit{\alpha}_{\mathcal{M}'}}\right)$ between vectors
$\boldit{\alpha}_{\mathcal{M}}$, $\boldit{\alpha}_{\mathcal{M}'}$,
\begin{equation}\label{18.2}
S_{\mathcal{M}\mathcal{M}'}^{\sigma}\equiv
\begin{cases}
\dfrac{\boldit{\alpha}_{\mathcal{M}}\cdot\boldit{\alpha}_{\mathcal{M}'}}{\alpha_{\mathcal{M}}\alpha_{\mathcal{M}'}}=
\cos\psi_{\mathcal{M}\mathcal{M}'}, \q \sigma=+,\\
\dfrac{\boldrm{e}_{z}\cdot(\boldit{\alpha}_{\mathcal{M}}\times\boldit{\alpha}_{\mathcal{M}'})}
{\alpha_{\mathcal{M}}\alpha_{\mathcal{M}'}}=
\sin\psi_{\mathcal{M}\mathcal{M}'},\q \sigma=- \, .
\end{cases}
\end{equation}

The resonances in the system are due to existence of the eigenmodes in the
film, i.e., SPPs. When an evanescent field harmonic is close to the grazing
one, its amplitude increases significantly as the process of eigenmode excitation
occurs. For infinitesimal modulation the eigenmodes of the film are initial
SPPs, with the dispersion relation corresponding to the determinant of the
matrix $\|D_{\mathcal{M}\mathcal{M}'}^{\tau\tau'|\sigma\sigma''}\|$ vanishing
in zeroth order approximation in modulation. Then
$\det\|D_{\mathcal{M}\mathcal{M}'}^{\tau\tau'|\sigma\sigma''}\|$ becomes an
infinite product of $\|b_{\mathcal{M}}^{\tau\tau'|\sigma}\|$ determinants, so
that each of them corresponds to some SPP eigenmode of the unmodulated film.
The equation $\det \|b_{\mathcal{M}}^{\tau\tau'|\sigma}\|=0$ has the physical
roots for $\sigma=-$ only, that corresponds to $p$ polarization:
\begin{eqnarray}\label{45.1}
\left(\beta_{+|\mathcal{M}}\tanh\Phi +
 \xi_{\mathcal{O}}\right)\left(\beta_{-|\mathcal{M}}\tanh\Phi +
 \xi_{\mathcal{O}}\right)\mathrm{cosh}^2\Phi \nonumber\\
 -\xi_{\mathcal{O}}^2 =0.
\end{eqnarray}
For a rather thick film, $\exp(\Phi') \gg 1$, SPPs existing in the film are close to those existing at the boundary
between the metal and each of the dielectric half-spaces, and are single-boundary-localized (SB) SPPs. These modes are
governed by the dispersion relation
\begin{equation}\label{45.2}
  \begin{split}
\beta_{\tau|\mathcal{M}} + \xi_{\mathcal{O}}=0,
\end{split}
\end{equation}
where $\tau=+(-)$ is for metal-substrate (superstrate) SB SPP. The symmetric surrounding (for instance, a free-standing
film) presents a specific case because of the coincidence of the Eq.~\eqref{45.2} solutions for different
$\tau$.\footnote{For a modulated film there can exist coupling between SB SPPs caused by diffraction. Such DB dressed
SPPs correspond to specific relations between permittivities of the dielectric media. Formally they correspond to
simultaneous vanishing of two determinants, $\|b_{\mathcal{M}}^{\tau\tau'|\sigma}\|$ and
$\|b_{\mathcal{M}'}^{\tau\tau'|\sigma}\|$, with $\mathcal{M}' \ne \mathcal{M}$, and in calculating the matrix
determinant $D_{\mathcal{M}\mathcal{M}'}^{\tau\tau'|\sigma\sigma''}$ (in the lowest order in the modulation) one has to
take into consideration the products of the determinants of the diagonal in diffraction orders submatrices
$b_{\mathcal{M}''}^{\tau\tau'|\sigma}$ along with the terms corresponding to the submatrices
$d_{\mathcal{M}\mathcal{M}'}^{\tau\tau'|\sigma\sigma'}$,
$d_{\mathcal{M}'\mathcal{M}}^{\tau\tau'|\sigma\sigma'}$.}\label{comment1} Then the initial SPPs existing at the
boundary of the metal and dielectric half-space become coupled due to the finite film thickness, and one obtains the
two double-boundary-localized (DB) SPP modes: long-range (LR) and short-range (SR)
SPPs\cite{LRSPP_PRB91,Surface-Polariton-Like_PRB86}. For $\varepsilon_+=\varepsilon_-\equiv\varepsilon$, and,
respectively, $\beta_{+|\mathcal{M}}=\beta_{-|\mathcal{M}} \equiv \beta_{\mathcal{M}}$,  one finds from
Eq.~\eqref{45.1} the two roots, $\beta_{\mathcal{M}} = \beta_{\mathcal{M}}^\pm$,
\begin{equation}\label{45.2.1}
  \begin{split}
\beta_{\mathcal{M}}^l=-\xi_{\mathcal{O}}\tanh(\Phi/2), \q \beta_{\mathcal{M}}^s=-\xi_{\mathcal{O}}\coth(\Phi/2) ,
\end{split}
\end{equation}
that is the single-boundary localized modes coupled into double-boundary localized ones. Therefore, the frequencies
of initial SPPs are repulsed, the spectral degeneration vanishes, and one arrives at two different eigenfrequencies
described by \equ{45.2.1}. Here superscript ``$l$'' (``$s$'') corresponds to the LR (SR) SPP. The LR (SR) mode possesses a high (low) frequency and is related to antisymmetric (symmetric) with respect to the midplane $z=d/2$ surface charge
distribution and the spatial distribution of the tangential to the film faces electric field component. As a result, SR
SPP possesses higher Ohmic losses.

Since the modulation is presumed to be small, the eigenmodes of the modulated
film (``dressed'' modes) differ slightly from those existing in the unmodulated
film. The dispersion relation of the dressed SPP modes defines the resonance
conditions. However for identification of the resonance type the modulation can be
neglected. Bearing in mind that $\beta_{\mathcal{M}|\tau}$ depends upon the
angle of incidence, $\theta$, and the wavelength, $\lambda$, it is possible to
consider the imaginary part of \eqref{45.2} as the ``resonance curve'' in
$\theta-\lambda$ plane. For instance, in the case of rectangular symmetry,
$\mathbf{g}_{1} \bot \mathbf{g}_{2}$, $g_{1} = g_{2}$, imaginary part of
\equ{45.2} reads
\begin{equation}\label{45.2.0}
 (\sin\theta\cos\psi+m_1\kappa_1)^2+(\sin\theta\sin\psi+m_2\kappa_2)^2=K_\tau^2,
\end{equation}
where $\psi$ is the angle of the incident plane orientation relative to $\mathbf{g}_1$ (say, ``tilting'' angle),
$\kappa_{1,2}=\mathrm{g}_{1,2}/k=\lambda/\rho_{1,2}$, and $K_\tau= \sqrt{\varepsilon_{\tau}+
\varepsilon_{\tau}^2{\xi_{\mathcal{O}}''}^2}$ is the dimensionless wavevector of SPP. We designate the curve given by
Eq.~\eqref{45.2.0} as $(m_1,m_2)_{\tau}$. All points of $\theta-\lambda$ plane may be classified as follows. Indeed,
the resonance curves possess the finite width due to the dissipative and radiative losses and, therefore, it is more
appropriate to refer to the resonance vicinity, but not to the resonance point. If a point does not belong to any
resonance curve, we have a nonresonance diffraction case. If a point belongs to a single curve with a fixed
$\mathcal{M}$ and $\tau$, we obtain a \emph{single diffraction-order single-boundary resonance (SSB)}. If a point
corresponds to intersection of several curves having different multiindexes
$\{\mathcal{M},\mathcal{M}',\mathcal{M}'',...\}$, but the same $\tau$ value, we obtain a \emph{multiple
diffraction-order single-boundary resonance}. Note that for specific geometry of high symmetry, $\psi=0$ and
$\psi=\pi/2$, the curves for different signs of $m_2$ and $m_1$ coincide; when $\psi=\pi/4$ curves $(m_1,m_2)_{\tau}$
and $(m_2,m_1)_{\tau}$ are indistinguishable as well. Intersection of two resonance curves with different $\tau$ yields
the point of a \emph{double-boundary resonance}. When DB SPPs have a unique multiindex $\mathcal{M}$  we then arrive at
a \emph{single diffraction-order double-boundary resonance (SDB)}. SDB resonance occurs only for the symmetric
surrounding of the film, $\varepsilon_+=\varepsilon_-$. Here the initial surface modes are coupled mainly through the
finite film thickness and, in general, the corresponding dependencies of the reflectance/transmittance are of
two-valley (two-peak) shape due to splitting of LR and SR modes. The explicit form of the resonance curve corresponding
to LR (SR) SPP resonances coincides with that given by Eq.~\eqref{45.2.0}, if one changes $\xi_{\mathcal{O}}''$ to
$\xi_{\mathcal{O}}''\tanh(\Phi'/2)$ [$\xi_{\mathcal{O}}''\coth(\Phi'/2)$] in the designation of $K_\tau$ in
Eq.~\eqref{45.2.0}. When DB SPPs are related to different multiindexes (\emph{multiple diffraction-order
double-boundary resonance}), which occurs under very specific conditions, they are coupled through periodicity and the
finite film thickness simultaneously.

Now take up the solution of the system \eqref{38}. In a small region of $\theta-\lambda$ plane, which is of order of
the resonance width, the solution is strongly dependent upon the number of resonance curves passing through this
region. Hence it is convenient to subdivide the set of the diffraction orders into a resonance subset, $\Re$, which
contains the multiindexes relative to the above-mentioned curves, and a nonresonance subset, $\mathfrak{N}$. Thus, we
subdivide the initial infinite system into the resonance subsystem containing the resonance TCs and the nonresonance
subsystem containing the nonresonance TCs. The resonance TCs correspond to $p$-components of the TCs with the resonance
multiindexes, and the nonresonance TCs correspond to $s$-components of amplitudes with the resonance multiindexes,
$T_{\mathcal{R}}^{\tau|+\sigma}$, and both $p$- and $s$-components of TCs with nonresonance multiindexes,
$T_{\mathcal{N}}^{\tau|\sigma\sigma'}$, $\mathcal{N} \in \mathfrak{N}$.

In the main approximation\cite{Ours_JETPHL_04,film_PhysRev,KNN_PhysRev}, which assumes
retaining the quadratic in modulation amplitude terms in the matrix elements, and the
linear terms in the right-hand sides, the resonance subsystem becomes:
\begin{equation}\label{r}
\sum_{\tau',\mathcal{R}'} B_{\mathcal{R}\mathcal{R}'}^{\tau\tau'} T^{\tau'|-\sigma}_{\mathcal{R}'}=
\tilde{V}_{\mathcal{R}}^{\tau|\sigma},
\end{equation}
where
\begin{eqnarray}\label{rs14}
&&\!\!\!\!\!\!\!
B_{\mathcal{R}\mathcal{R}'}^{\tau\tau'}=\delta_{\mathcal{R}\mathcal{R}'}
b_{\mathcal{R}}^{\tau\tau'|-}+
d_{\mathcal{R}\mathcal{R}'}^{\tau\tau'|--}\nonumber\\
&&\!\!\!\!\!\!\!-\overline{\sum_{\mathcal{M},\sigma''}}
\sum_{\tau'',\tau'''}d_{\mathcal{R}\mathcal{M}}^{\tau\tau''|-\sigma''}
\left(\widehat{b}^{-1}\right)_{\mathcal{M}}^{\tau''\tau'''|\sigma''}
d_{\mathcal{M}\mathcal{R}'}^{\tau'''\tau'|\sigma''-},
\end{eqnarray}
\begin{eqnarray}\label{rs9.1}
\tilde{V}_{\mathcal{R}}^{\tau|\sigma}=V_{\mathcal{R}}^{\tau|\sigma\sigma'}\!\!- \overline{\delta}_{\Re
\mathcal{O}}^{\frac{1-\sigma}{2}} \sum_{\tau',\tau''}d_{\mathcal{R}\mathcal{O}}^{\tau\tau'|-\sigma}
\left(\widehat{b}^{-1}\right)_{\mathcal{O}}^{\tau'\tau''|\sigma} V_{\mathcal{O}}^{\tau''|\sigma}\!\!.
\end{eqnarray}
Here $\widehat{b}\equiv\|b_{\mathcal{M}}^{\tau\tau'|\sigma}\|$, the sum with the overline, $\overline{\sum}$, means
that the terms with the superscript $\sigma''=-$ and a resonance diffraction order $\mathcal{M} = \mathcal{R}\in\Re$
have to be omitted. The function $\overline{\delta}_{\Re \mathcal{O}}$ is equal to $1$, if within the resonance indexes
there is the zeroth one, and to $0$ otherwise. The zero-value of matrix $\widehat{B}$ determinant yields a dispersion
relation of the SPP modes in the film in the main approximation. As was discussed above, the block,
$b_{\mathcal{R}}^{\tau\tau'|-}$, being diagonal relative to the diffraction order, contributes to the unperturbed
dispersion relation. The nondiagonal, linear-in-modulation block, $d_{\mathcal{R}\mathcal{R}'}^{\tau\tau'|--}$,
contains the ``interresonance'' or ``coupling'' modulation harmonic, $\tilde{\xi}_{\mathcal{R}-\mathcal{R}'}$, which is
chiefly responsible for the SPP dispersion curve splitting, and for the appearance of the spectral band
gap\cite{SPIE,Maradud_2D_gaps_PRB02}. The third term in \eqref{rs14} is quadratic in modulation, and describes the
second-order scattering processes which provide the broadening and the shift of the dispersion branches.

%The elements of the last block in \eqref{rs14} contain the quadratic terms describing the second-order
%scattering processes which provide the broadening and the shift of the dispersion branches.

The nonresonance TCs, $T_{\mathcal{N}}^{\tau|\sigma\sigma'}$, are expressed in terms of the resonance
TCs as
\begin{equation}\label{r1}
T^{\tau|\sigma\sigma'}_{\mathcal{N}}\simeq
\delta_{\mathcal{N},\mathcal{O}}\delta_{\sigma\sigma'}T_{F}^{\tau|\sigma}-
\sum_{\tau',\tau''}\left(\widehat{b}^{-1}\right)_{\mathcal{N}}^{\tau\tau'|\sigma}
d_{\mathcal{N}\mathcal{R}}^{\tau'\tau''|\sigma-} T_{\mathcal{R}}^{\tau''|-\sigma'}\!\!,
\end{equation}
$T_{F}^{\tau|\sigma}$ is a transmission (for $\tau=+$) or a reflection (for $\tau=-$) coefficient corresponding to $p$ (for $\sigma=-$) or $s$ (for $\sigma=+$) polarization in the case of an unmodulated film. $s$-components of the TCs with the resonance multiindexes, $T_{\mathcal{R}}^{\tau|+\sigma}$,
are expressed in terms of $p$-components, $T_{\mathcal{R}}^{\tau|-\sigma}$, and are small as compared with the latter; they are of no interest, and we do not present them herein. The diffraction efficiencies are %(giving a ratio of the $z$-
%components of the Poynting vectors corresponding to the homogeneous waves
%within the dielectrics)
$$
\tau_{\mathcal{M}}=\frac{\left|\boldrm{E}_{\mathcal{M}}^{+}\right|^2}
{\left|\boldrm{E}\right|^2}\cdot\frac{\Re e(k_{+|\mathcal{M}z})}{k_{z}}, \quad
\rho_{\mathcal{M}}=\frac{\left|\boldrm{E}_{\mathcal{M}}^{-}\right|^2}
{\left|\boldrm{E}\right|^2}\cdot \frac{\Re e(k_{-|\mathcal{M}z})}{k_{z}} .
$$
From the energy conservation it evidently follows that
\begin{equation}\label{4_10_05}
 1 - \sum_{\mathcal{M}} [\rho_{\mathcal{M}} + \tau_{\mathcal{M}}]  = \mathcal{P} \ge 0 ,
\end{equation}
where $\mathcal{P}\sim\int\varepsilon''|\overline{\mathbf{E}}|^2dz$ is the
absorbed part of the energy flux density.

\subsection{\label{subsec:single} Single diffraction-order resonance}

If a sertain point of $\lambda-\theta$  plane belongs to one or two curves \eqref{45.2.0} having a single multiindex
$\mathcal{R}=(r_1,r_2)$ (for other parameters being fixed) the vicinity of this point defines a single
diffraction-order resonance. Moreover, according to the above classification, in the case of one curve (for a single
$\tau$), we have a SB resonance, and if the point is the intersection of two curves with different $\tau$ values, we
have a DB resonance. The resonance TCs are then similar to those obtained for 1D modulation, cf.
Ref.~\onlinecite{KNN_PhysRev},
\begin{equation}\label{ss1}
   \left[ \!
\begin{array}{cc}
  T^{\tau|-+}_{\mathcal{R}} \\
  T^{\tau|--}_{\mathcal{R}} \\
\end{array}
\! \! \right]=
 \left[ \! \!
\begin{array}{cc}
  -\cos\theta\sin\psi_{\mathcal{R}\mathcal{O}} \\
  \cos\psi_{\mathcal{R}\mathcal{O}} \\
\end{array}
\! \! \right] L^{\tau}_{\mathcal{R}}\tilde{\xi}_{\mathcal{R}} [\exp(-\Phi)]^{\frac{1+\tau}{2}},
\end{equation}
where
\begin{equation}\label{ss2}
L_\mathcal{R}^{\tau}=2\tau (\tilde{\beta}_{\overline{\tau}|\mathcal{R}}-\delta_{\tau,+}
\Upsilon_\mathcal{R})\left/\Delta_{\mathcal{R}}\right.  ,
\end{equation}
\begin{equation}\label{ss3}
 \tilde{\beta}_{\tau|\mathcal{R}}=\beta_{\tau|\mathcal{R}}\tanh\Phi +
\xi_{\mathcal{O}}  + G_{\mathcal{R}}^{\tau}
 \, ,
\end{equation}
\begin{equation}\label{ss4}
\Delta_{\mathcal{R}}=\tilde{\beta}_{+|\mathcal{R}}\tilde{\beta}_{-|\mathcal{R}}-
\Upsilon_\mathcal{R}^{2}\mathrm{cosh}^{-2}\Phi, \;
 \Upsilon_\mathcal{R}=\xi_{\mathcal{O}} +G_{\mathcal{R}}^{+}+G_{\mathcal{R}}^{-}
 \; ,
\end{equation}
\begin{equation}\label{ss5}
G_{\mathcal{R}}^{\tau}=-\sum_{\mathcal{N}} \!\frac{\tilde{\xi}_{\mathcal{R}-\mathcal{N}}
\tilde{\xi}_{\mathcal{N}-\mathcal{R}}}{\beta_{\tau|\mathcal{N}}}
\!\left(\cos^2\psi_{\mathcal{R}\mathcal{N}}+\varepsilon_{\tau}\beta^2_{\tau|\mathcal{N}}\sin^2
\psi_{\mathcal{R}\mathcal{N}}\right)\!\!.
\end{equation}
Stress that $T_{\mathcal{R}}^{\tau|+\sigma}\sim O(\tilde{\xi}_\mathcal{R})$,
$|T_{\mathcal{R}}^{\tau|-\sigma}|\gg|T_{\mathcal{R}}^{\tau|+\sigma}|$. The resonance TCs have two poles. For a
nonsymmetric surrounding, these poles are releted to SPPs existing at the opposite film faces, while in the case of a
symmetric surrounding, they are related to LR and SR SPP modes.

The structure of the resonance TCs indicates that the coupling strength between the incident wave and the SPP excited
is proportional to the scalar product of the SPP and the incident wave magnetic fields. Note that the SPP magnetic
field is orthogonal to its propagation direction and is parallel to the interface so that the SPP is effectively
excited by the projection of the tangential component of incident wave magnetic field onto
$\mathbf{H}_{\mathcal{R}t}^\tau$ (or, alternatively, by the scalar product of tangential components of the electric SPP
fields and the incident wave). As for $p$($s$)-polarization of the incident light, the projection is
$H_t\cos\psi_{\mathcal{R}\mathcal{O}}$ ($H_t\cos\theta\sin\psi_{\mathcal{R}\mathcal{O}}$), where
$\psi_{\mathcal{R}\mathcal{O}}$ is the angle between $\mathbf{H}_{\mathcal{R}t}^\tau$ and $\mathbf{H}_t$. For instance,
with the incidence of purely $p$- ($s$-) polarized light, the SPPs propagating parallel (perpendicular) to the
incidence plane cannot be excited, even if wavevector $\boldrm{k}_{\mathcal{R}t}$ is close to the poles in \equ{ss1}.

The zeroth-order polarization matrix is
\begin{eqnarray}\label{ss6}
&&\!\!\!\!\begin{pmatrix}
T_{\mathcal{O}}^{\tau|++} & T_{\mathcal{O}}^{\tau|+-}\\
T_{\mathcal{O}}^{\tau|-+} & T_{\mathcal{O}}^{\tau|--}
\end{pmatrix}=
\begin{pmatrix}
T_{F}^{\tau|+}& 0\\
0 & T_{F}^{\tau|-}
\end{pmatrix}\nonumber\\
&&\!\!\!\!+ \!\!
\begin{pmatrix}
2\cos\theta_{\tau}\cos\theta\sin^2\psi_{\mathcal{R}\mathcal{O}}
 & \!-\cos\theta_{\tau}\sin2\psi_{\mathcal{R}\mathcal{O}}\\
-\tau\cos\theta\sin2\psi_{\mathcal{R}\mathcal{O}} &
\!2\tau\cos^2\psi_{\mathcal{R}\mathcal{O}}
\end{pmatrix}
 L_{\mathcal{O}|\mathcal{R}}^{\tau},\q\;\;
\end{eqnarray}
where
\begin{eqnarray}\label{ss7}
L_{\mathcal{O}|\mathcal{R}}^{\tau}=-\dfrac{\sqrt{\varepsilon_{\tau}}
\tilde{\xi}_{\mathcal{R}}\tilde{\xi}_{-\mathcal{R}}}{\cos\theta_{\tau} \Delta_{\mathcal{R}}}
\left[\tilde{\beta}_{\overline{\tau}|\mathcal{R}}+ \left(\tilde{\beta}_{\tau|\mathcal{R}}-
\Upsilon_\mathcal{R}\right)(\mathrm{cosh}\Phi)^{\tau-1}
\right]\nonumber\\
\!\!\!\times(\mathrm{cosh}\Phi)^{-\frac{1+\tau}{2}}.\q
\end{eqnarray}
$\theta_{\tau}$ is the angle of propagation of the zeroth-order wave in $\tau$th dielectric media relative to $Oz$ axis
(in the superstrate $\theta_{-}\equiv\theta$). The structure of these coefficients shows an interference caused by the
competition between the nonresonance channel (the terms $T_{F}^{\tau|\sigma}$), and the resonance channel (the second
terms in \equ{ss6}, which is proportional to
$\tilde{\xi}_{\mathcal{R}}\tilde{\xi}_{-\mathcal{R}}/\Delta_{\mathcal{R}}$). The nonresonance term should be retained
for the zeroth-order reflectance, since it is of order of unity, $|T_{F}^{-|\sigma}|\sim1$, while for a rather strong
resonance it can be neglected for the transmittance in the vicinity of the resonance maxima, since
$|T_{F}^{+|\sigma}|\sim|\xi_{\mathcal{O}}|\exp(-\Phi')\ll 1$ and is much smaller than the resonance input into the
transmittance.

The wavelength resonance width, $\Delta\lambda/\lambda$, is contributed both from the dissipation losses, being
proportional to $\xi_{\mathcal{O}}'$, and from the radiation losses due to SPP scattering into the outgoing propagating
waves. It is of order $\Delta\lambda/\lambda\sim|\xi_{\mathcal{O}}''|[\xi_{\mathcal{O}}'+O(|\tilde{\xi}^2|)]$, as it
follows from \eqref{ss1}-\eqref{ss7}. The term proportional to $O(|\tilde{\xi}^2|)$ is caused by the radiation losses
and in the simplest case may be represented approximately as
$\sum_{\mathcal{N}}C_{\mathcal{N}}\tilde{\xi}_{\mathcal{R}-\mathcal{N}}
\tilde{\xi}_{\mathcal{N}-\mathcal{R}}/\beta_{\tau|\mathcal{N}}$, where $|C_{\mathcal{N}}|\sim1$, and the summation is
performed over those $\mathcal{N}$ that satisfy $\Im m (\beta_{\tau|\mathcal{N}})=0$. According to the formulas
describing the resonance and zeroth-order TCs, the \emph{optimal amplitude} of the resonance harmonic is
$|\tilde{\xi}_{\mathcal{R}}|\sim\sqrt{\xi_{\mathcal{O}}'}$ and this leads to the resonance width of order
$\Delta\lambda/\lambda\sim \xi_{\mathcal{O}}'|\xi_{\mathcal{O}}''|$. The optimal amplitude is related to the maximal
SPP excitation and, consequently, to minimal/maximal value of reflectance/transmittance. In spite of the fact that we
assume the modulation to be rather small, let us make a rough astimation of the resonance width for the hole arrays.
For Ag films in the visible and near infra-red frequency region the impedance of the film is $|\xi_{f}|\sim10^{-1}$,
while for the holes the impedance $|\xi_{i}|\sim1$. Therefore,
$|\tilde{\xi}_{\mathcal{R}}|\sim|\Delta\xi|\sim1\gg\xi_{\mathcal{O}}'$, and
$\Delta\lambda/\lambda\sim|\xi_{\mathcal{O}}''|\sim10^{-1}$. For wavelength of order $\lambda\sim1\mu m$ the resonance
width can be estimated as $\Delta\lambda\sim100nm$, which is in good qualitative agreement with numerous experimental
results. In the experiments the resonance width is equally affected by the non-plane character of an input light wave
and the finiteness of the periodic array.

Note that the resonance width is very important for problems of
sub-diffraction-limited optical imaging (optics of volume or surface
``superlenses''). Sufficiently broadened SPP resonances may be efficiently
used to enhance the evanescent modes and thus to recover the
subwavelength information on nanoobjects, see
Ref.~\onlinecite{superlens_science05}.

\subsection{\label{subsec:multiple}Multiple resonances}

The approach developed allows one to consider the diffraction problem for the resonances of arbitrary multiplicity on
2D-periodical structures possessing an arbitrary symmetry. But from the experimental point of view the resonances of
fourfold multiplicity, which take place for the normal incidence, $\theta = 0$, at the square and rectangular
periodical arrays are of special interest. Here, and in what follows, we will concentrate on the structures possessing
$C_{2v}$ symmetry: $\tilde{\xi}(\widehat{C}_{2v}\mathbf{r}_{t})=\tilde{\xi}(\mathbf{r}_{t})$; the geometrical symmetry
(Brillouin zone symmetry) is supposed to be $C_{4v}$, i.e. $\mathbf{g}_1$ is perpendicular to $\mathbf{g}_2$ and their
modules are equal, $\mathrm{g}_1=\mathrm{g}_2$. As it seen from Eq.~\eqref{45.2.0}, at normal incidence the multiple
SPP resonances arise at the wavelengths $\lambda_{r_1,r_2}^\tau = \rho K_\tau/\sqrt{r_1^2+r_2^2}$, where $\rho$ is the
period of the structure. Points $\theta = 0$, $\lambda=\lambda_{r_1,r_2}^\tau$ in the $\theta-\lambda$ plane are the
intersections of four resonance curves for the single $\tau$ (SB resonance) or eight curves for both $\tau=\pm$ (DB
resonance in symmetric surrounding). We mark a fourfold SB resonance as $[r_1,r_2]_{\tau}$, which corresponds to the
intersection of the following resonance curves: $(r_1,r_2)_{\tau}$, $(-r_1,r_2)_{\tau}$, $(r_1,-r_2)_{\tau}$, and
$(-r_1,-r_2)_{\tau}$. A fourfold DB resonance is marked as $[r_1,r_2]$ respectively. In addition, if all values of a
certain function $F_{\mathcal{M}}$ with subscripts from the above subsets are equal, we denote it by $F_{[r_1,r_2]}$.

Similarly to the single resonance, when we have the fourfold one, the pair of field harmonics with wavevectors
$\boldrm{k}_{(r_1,r_2)t}=r_1\mathbf{g}_1+r_2\mathbf{g}_2$ and
$\boldrm{k}_{(-r_1,-r_2)t}=-r_1\mathbf{g}_1-r_2\mathbf{g}_2$ is efficiently generated by the projection of the incident
wave magnetic field onto the direction perpendicular to $\boldrm{k}_{(r_1,r_2)t}$ (the direction, corresponding to the
SPP magnetic field with this wavevector), viz. by $H\sin\phi_{r_1,r_2}$, where $\phi_{\mathcal{R}}$ is the angle
between $\boldrm{k}_{\mathcal{R}t}$ and $\mathbf{H}_t = \mathbf{H}$. Analogously, the amplitudes of the field harmonics
corresponding to diffraction indexes $(\pm r_1,\mp r_2)$ are proportional to $H\sin\phi_{-r_1,r_2}$. Thus, for special
polarization of the incident wave, at $\phi_{\mathcal{R}}=0,\pm\pi$, $\pm\mathcal{R}$th the resonance field harmonics
are not excited via the first-order scattering process (but rather via the higher-order processes). This reduces the
multiplicity of the resonance twice in the main approximation. For an arbitrary polarized incident wave, the
polarizations of the zeroth-order transmitted and reflected waves are formed mainly by the interference contribution of
all resonance field harmonics due to single back-scattering. Also, they are contributed from the zeroth-order
components related to the ``scatteringless'' reflection and transmission for an unmodulated film. Otherwise, the
polarizations of both propagating and evanescent nonresonance field harmonics are formed mainly by the single
scattering from the zeroth diffraction order and from all the resonance diffraction orders.

For the structures possessing $C_{2v}$ modulation symmetry, the solution of \eqref{r} is considerably simplified in the
vicinity of normal incidence. In the case of $[r,0]$ or $[r,0]_\tau$ resonance, the resonance TCs are similar to those
given in \equ{ss1}, cf. explicit expressions in Appendix \ref{C2v}, Eq.~\eqref{b1}. Though in
$\tilde{\beta}_{\tau|\mathcal{R}}$ a linear-in-modulation term, $\tilde{\xi}_{2\mathcal{R}}$, arises, which is the
inter-resonance harmonic responsible for the splitting and shifting of the resonance.

To get a better understanding of the resonance diffraction, consider the eigenmodes of $C_{2v}$ structures. They are defined approximately by relation $\widehat{B}\widehat{T}=0$. In particular, the eigenfrequencies may be found from
equation $\det\widehat{B}=0$. For SB-localized SPPs when we have a rather thick film, $\exp(-\Phi')\ll1$, the
structure of eigenmodes can be obtained in the approximation of the half-space problem, see the detailed analytical
treatment in Ref.~\onlinecite{SPIE}. Thus, we restrict ourselves to the eigenmodes that are close to those
corresponding to the metal-dielectric interface and not coupled through film thickness. Bearing in mind the
homogeneous problem setup, we must change the notations so that one of the resonance $\mathbf{k}$-vectors in
 the diffraction problem, $\mathbf{k}_{\mathcal{R}t}$, become the SPP quasi-wavevector, $\mathbf{q}$, ending in some
Brillouin zone. The other wavevectors close to the resonance conditions correspond to the ``resonance satellites'',
that compose a coupled SPP state.

\begin{figure}
  \includegraphics[width=6cm]{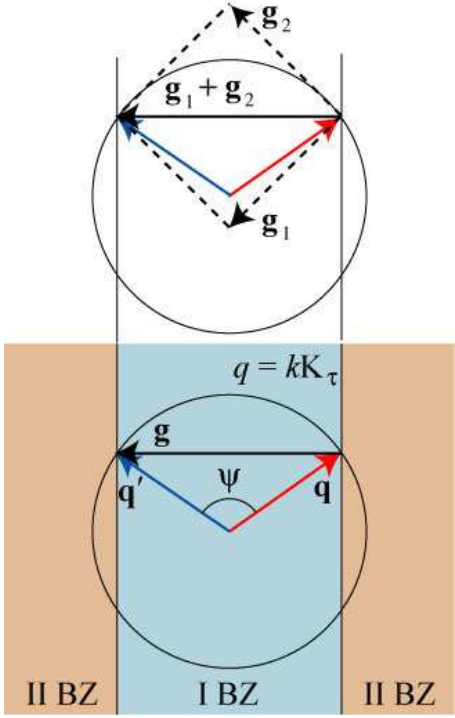}\\
  \caption{\label{BZ}(Color online) An example of a twofold coupling of the SPPs in the
case where $\Delta \mathcal{R} = (1,1)$. The upper figure is for the $2D$ problem, and the lower is for the similar
$1D$ one.}
\end{figure}

Concentrate first on the simplest case of a twofold coupling through the periodicity. Suppose that the resonance
wavevectors are $\mathbf{k}_{{R}t}\rightarrow \mathbf{q}$ and $\mathbf{k}_{{R}'t}\rightarrow \mathbf{q}'$. Then we can
consider the diagonal in $\tau$ homogeneous subsystems of the resonance system \eqref{r} which contains the two TCs,
for the diffraction orders $\mathcal{R}=(r_1,r_2)$ and $\mathcal{R}'=(r'_1,r'_2)$. Using
Eqs.~\eqref{44}-\eqref{18.2} (and assuming $\tanh\Phi=1$), we write it as
\begin{equation}\label{mr0}
 \begin{split}
\begin{pmatrix}
\beta_{\tau|\mathcal{R}}+\xi_{\mathcal{O}} & \tilde{\xi}_{\Delta \mathcal{R}}
\cos\psi_{\mathcal{R}\mathcal{R}'}\\
\tilde{\xi}_{-\Delta \mathcal{R}}\cos\psi_{\mathcal{R}\mathcal{R}'} & \beta_{\tau|\mathcal{R}'}+\xi_{\mathcal{O}}
\end{pmatrix}
\begin{pmatrix}
E_{\mathcal{R}z}^{\tau}\\
E_{\mathcal{R}'z}^{\tau}
\end{pmatrix}=0.
 \end{split}
\end{equation}
Here $\Delta \mathcal{R} = \mathcal{R} - \mathcal{R}'$, and we neglect the quadratic-in-modulation amplitude terms,
supposing that they do not exceed the linear-in-coupling harmonic $\tilde{\xi}_{\pm\Delta \mathcal{R}}$ term. We take
into account that in the $\tau$th dielectric half-space the SPP electric field has predominantly a $z$-component and
therefore, change $T^{\tau|-\sigma}_{\mathcal{R}}$ to $E_{\mathcal{R}z}^{\tau}$. From this system we see immediately
that the initial SPPs with wave vectors oriented at $\psi_{\mathcal{R}\mathcal{R}'}=\pm\pi/2$ are not coupled, and it
is thus the case where it is necessary to retain the quadratic coupling terms\footnote{In case of the periodicity
formed by the surface relief corrugation, the situation changes principally: coupling does exist even when SPPs
propagate perpendicular to one another.}. Note that, as we see, here the $2D$ problem is reduced to the $1D$ one,
corresponding to the twofold SPP coupling through the harmonic grating of the period
$2\pi/|\mathbf{k}_{{R}_{1}t}-\mathbf{k}_{{R}_{2}t}|$ (and with amplitude $\tilde{\xi}_{\pm\Delta \mathcal{R}}$), where
the quasi-wavevector of the dressed SPP, $\mathbf{q}$, and the satellite, $\mathbf{q}' = \mathbf{q} +
(r'_1-r_1)\mathbf{g}_1+ (r'_2-r_2)\mathbf{g}_2$, refer to the sides of a Brillouin zone, see Fig.~\ref{BZ}. Then
$\beta_{\tau|\mathcal{R}}=\beta_{\tau|\mathcal{R}'}\equiv\beta_{\tau}$, where
$\beta_{\tau}=\sqrt{\varepsilon_{\tau}-(qc/\omega)^{2}}/\varepsilon_{\tau}$, and the dispersion relation becomes

\begin{equation}\label{mr1}
\beta_{\tau}(\omega^{\pm}) = -\xi_{\mathcal{O}} \mp \sqrt{\tilde{\xi}_{\Delta r}\tilde{\xi}_{-\Delta r}} \, \c |
\cos\psi_{ \mathcal{R} \mathcal{R}'}|.
\end{equation}
Remind that $\psi_{ \mathcal{R} \mathcal{R}'}$ is the angle between $\mathbf{q}$ and $\mathbf{q}'$. Hence, we obtain
the two roots: the first one is for the high frequency mode, $\omega^{+}$, and the second one is for the low-frequency
mode, $\omega^{-}$. Specifically, neglecting the modulation of the small real part of the surface impedance as compared
with that of its imaginary part we have $\tilde{\xi}_{-m} = - \tilde{\xi}_{m}^\ast$, and $\sqrt{\tilde{\xi}_{\Delta
\mathcal{R}}\tilde{\xi}_{-\Delta \mathcal{R}}} = i |\tilde{\xi}_{\Delta \mathcal{R}}|$. Hence, the eigenfrequencies are
\begin{eqnarray}\label{mr1.1}
\omega^{\pm}/\omega_{ph} \simeq 1+ \varepsilon_\tau\xi_{\mathcal{O}}^2/2 \pm i\varepsilon_\tau\xi_{\mathcal{O}}
|\tilde{\xi}_{\Delta \mathcal{R}}
\cos\psi_{ \mathcal{R} \mathcal{R}'}|, \nonumber\\
\omega_{ph}=qc/\sqrt{\varepsilon_\tau} \; ,
\end{eqnarray}
where the upper (lower) sign is for high (low) frequency SPP. It is important to note that the quality($Q$)-factor for
the high-frequency mode is higher than that for the low-frequency mode.

\begin{figure}[!tb]
  \includegraphics[width=6cm]{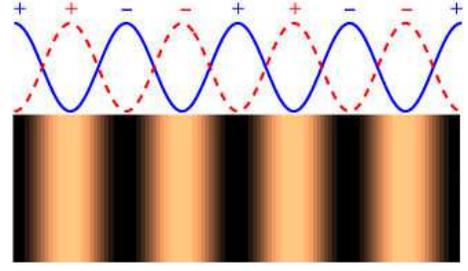}\\
  \caption{\label{grat}(Color online) Spatial distribution of the squared electric field
$z$-components of the eigenmodes. The solid (dashed) curve is for the low (high)- frequency mode. The ``more metallic''
regions of the coupling harmonic contribution are shown by the darker regions of the contourplot. Pluses and minuses
refer to the signs of the surface charge density.}
\end{figure}

For the specific choice of coordinate origin, such that $\tilde{\xi}_{\Delta \mathcal{R}} = i |\tilde{\xi}_{\Delta
\mathcal{R}}|$, $\xi(\mathbf{r}_{t})=\xi_{\mathcal{O}}+...+ 2i|\tilde{\xi}_{\Delta \mathcal{R}}|
\cos[(\mathbf{q}-\mathbf{q}')\mathbf{r}]+...$, the electric field amplitudes of the eigenmodes obey the relation
$\left(E_{z \mathcal{R}}^{\tau}/E_{z \mathcal{R}'}^{\tau}\right)^\pm=\pm \mathrm{sign}[\cos\psi_{ \mathcal{R}
\mathcal{R}'}]$. As the coupled SPPs propagate in the opposite directions, $\psi=\pm\pi$, we have
$\mathbf{q}=-\mathbf{q}'$, and the field structure at the interface for the eigenmodes is
\begin{equation}\label{mr2}
E_z^{\tau(-)}(\mathbf{r}_t)\sim\cos(\mathbf{q}\mathbf{r}), \q
E_z^{\tau(+)}(\mathbf{r}_t)\sim\sin(\mathbf{q}\mathbf{r}),
\end{equation}
Note that under the condition of the prevalence of the coupling in the first order scattering, or, in other words, when
the coupling harmonic dominates over the others, $|\tilde{\xi}_{\Delta \mathcal{R}}|>|\tilde{\xi}_{\mathcal{M}}|^2$,
the structure of the spatial field distribution is given by the coupling harmonic. In other words, the field maxima of
the high frequency mode coincide with the ``less metallic'' regions (where $|\xi|$ is higher) relative to the coupling
harmonic and vice versa for the low frequency mode, see Fig.~\ref{grat}.

Discuss the eigenmodes for fourfold SB SPP resonance. We limit ourselves to $[r,0]$ resonance, such that
$\boldrm{k}_{(r,0)t}=-\boldrm{k}_{(-r,0)t}$, $\boldrm{k}_{(0,r)t}=-\boldrm{k}_{(0,-r)t}$, and $\boldrm{k}_{(0,\pm r)t}$
$\bot$ $\boldrm{k}_{(\pm r,0)t}$. Remind that the resonance amplitudes corresponding to perpendicularly oriented
vectors are not coupled in the first scattering order. One can make sure that four eigenmodes exist. Two of them are
``mixed'', that is, all resonance amplitudes are nonzero: $E_{(\pm r,0)z}^{\tau},E_{(0,\pm r)z}^{\tau} \neq0$,
moreover, $E_{(r,0)z}^{\tau}=E_{(-r,0)z}^{\tau}$, and $E_{(0,r)z}^{\tau}=E_{(0,-r)z}^{\tau}$. Thus, the field structure
has the form of the linear combination of two cosines having their periods along $\mathbf{g}_1$ and $\mathbf{g}_2$.
These modes cannot be excited at
 normal incidence. Other two modes, which may be excited, have zero resonance amplitudes: for one mode
$E_{(r,0)z}^{\tau}=-E_{(-r,0)z}^{\tau}$, $E_{(0,\pm r)z}^{\tau} =0$, and $E_{(0,r)z}^{\tau}=-E_{(0,-r)z}^{\tau}$,
$E_{(\pm r,0)z}^{\tau} =0$ for another one. Consequently, the spatial structure corresponds to sinuses. Thus, the
structure of the eigenmodes in this particular instance of fourfold SB SPP resonance is similar to that of twofold SB SPP
resonances; they have the form of a pair of standing SPPs, that correspond to the sinus-type spatial field distribution, and higher-frequency (and also higher Q-factor) branches. This means that in Eqs.~\eqref{mr1.1} taking ``$+$'' we should
set $|\cos\psi_{ \mathcal{R} \mathcal{R}'}|\rightarrow1$, $\Delta r\rightarrow(2r,0)$, and the first (second) standing
SPP quasi-wavevector, $\mathbf{q}$, becomes parallel to $\boldrm{k}_{(r,0)t}$ [$\boldrm{k}_{(0,r)t}$].

The solution of the inhomogeneous problem is the superposition of the above standing SPPs with the ``weights''
proportional to cosines of the angles between $\mathbf{H}_{(r,0)t}^\tau$, $\mathbf{H}_{(0,r)t}^\tau$, and $\mathbf{H}$.
The eigenmodes for other SB-fourfold resonances may be treated in a similar way, they are the combinations of simplest
SPP modes as well.

For DB-localized dressed SPPs the field structure is defined by coupling of the initial SPPs both through the
modulation and through the finite film thickness. The field structure resulting from the coupling through the film
thickness may be understood from the example of the undressed SPP existing in the unmodulated film. The
eigenfrequencies for this case are defined by Eq.~\eqref{45.2.1}. Note that inside the metal the amplitude of the
tangential-to-interface component of the electric field is higher than $z$-component,
$|\overline{E}_t/\overline{E}_z|\sim|\sqrt{\varepsilon}|\gg1$, in contrast to the dielectric half-spaces. Remind also
that $z$-dependence of the electric field inside the film for LR and SR eigenmodes
is\cite{Book_Raether,Book_Agranovich}
\begin{equation}\label{mr3}
\overline{E}_t^{(L)}\sim\sinh(\tilde{k}z-\Phi/2), \q \overline{E}_t^{(S)}\sim\cosh(\tilde{k}z-\Phi/2),
\end{equation}
where $(L)$ stands for LR SPP and $(S)$ for SR SPP.

For dressed DB-localized SPPs with the twofold coupling, the dispersion relation is
\begin{eqnarray}\label{mr4}
\beta^{(L\pm)}= -(\xi_{\mathcal{O}}\pm\tilde{\xi}_{\Delta \mathcal{R}}|
\cos\psi|)\tanh(\Phi/2),\nonumber \\
\beta^{(S\pm)}= -(\xi_{\mathcal{O}}\pm\tilde{\xi}_{\Delta \mathcal{R}}| \cos\psi|)\coth(\Phi/2),
\end{eqnarray}
where $\beta^{(...)}=\sqrt{\varepsilon_{d}-(qc/\omega)^{2}}/\varepsilon_{d}$, $\varepsilon_+=
\varepsilon_-=\varepsilon_d$, superscripts in brackets mark the eigenmodes. Here the initial SPPs (and corresponding
diffraction indexes) are the same as in the above case of the dressed twofold SB SPP. The spatial field
distribution for eigenmodes inside the film for $\psi=\pi$ ($\mathbf{q}=-\mathbf{q}'$) has the form
\begin{eqnarray}\label{mr5}
\overline{\mathbf{E}}_t^{(S-)}(\mathbf{r})\sim \hat{\mathbf{e}}_{\mathbf{q}}\cosh(\tilde{k}z-\Phi/2)
\sin(\boldrm{q}\mathbf{r}),
\nonumber\\
\overline{\mathbf{E}}_t^{(L-)}(\mathbf{r})\sim \hat{\mathbf{e}}_{\mathbf{q}}\sinh(\tilde{k}z-\Phi/2)
\sin(\boldrm{q}\mathbf{r}),
\nonumber\\
\overline{\mathbf{E}}_t^{(S+)}(\mathbf{r})\sim \hat{\mathbf{e}}_{\mathbf{q}}\cosh(\tilde{k}z-\Phi/2)
\cos(\boldrm{q}\mathbf{r}),
\nonumber\\
\overline{\mathbf{E}}_t^{(L+)}(\mathbf{r})\sim \hat{\mathbf{e}}_{\mathbf{q}}\sinh(\tilde{k}z-\Phi/2)
\cos(\boldrm{q}\mathbf{r}).
 \end{eqnarray}
$z$-components of the electric field in the dielectrics at the film faces are of the same form as in \equ{mr2}, i.e.
$E_z^{\tau(S-)}(\mathbf{r})\sim E_z^{\tau(L-)}(\mathbf{r})\sim\cos(\mathbf{q}\mathbf{r})$ and
$E_z^{\tau(S+)}(\mathbf{r})\sim E_z^{\tau(L+)}(\mathbf{r})\sim\sin(\mathbf{q}\mathbf{r})$. It is possible to obtain the
eigenfrequencies and the field structure for the fourfold DB SPP, proceeding in the same way, in which we generalized
the twofold SB SPP to the fourfold SB SPP.

Analyzing the denominator of TCs, on can make sure that the excited DB SPPs are related to ``$+$'' modes only. Indeed,
with the film thickness tending to infinity, we can see that the two poles of TCs in Eq.~\eqref{b1} of
Appendix~\ref{C2v}, $\Delta_{\mathcal{R}}=0$, become $\beta_{\tau|\mathcal{R}} +
\xi_{\mathcal{O}}+\tilde{\xi}_{2\mathcal{R}} =0$, in neglecting the second-order scattering processes. Choosing an
origin such that $\tilde{\xi}_{2\mathcal{R}}=i|\tilde{\xi}_{2\mathcal{R}}|$, the poles coincides with ``$+$'' branch of
Eq.~\eqref{mr1}, where we have $\tilde{\xi}_{2\mathcal{R}}$ instead of $\tilde{\xi}_{\Delta\mathcal{R}}$, and $|
\cos\psi_{ \mathcal{R} \mathcal{R}'}|=1$. This corresponds to the upper frequency branch, Eq.~\eqref{mr1.1}, and,
hence, to the ``sine'' field distribution in the $x-y$ plane.

Under the assumption of $C_{4v}$ modulation symmetry the zeroth-order TCs for SB $[r,0]_{\tau}$ resonance have similar
to \equ{ss6} structure as well (see Appendix \ref{C2v}). However, the difference consists in identity
$T^{\tau|\sigma\overline{\sigma}}_{\mathcal{O}}=0$, which ensures that the polarization of the zeroth-order transmitted
wave coincides with that of the incident wave.

Along with the anomalies related to SPP excitation there exist Rayleigh anomalies. They correspond to the boundary
between homogeneous (propagating) and inhomogeneous (evanescent) waves in different diffraction orders, viz to the
vanishing of $z$-component of one of the wave-vectors, $k_{\tau|\mathcal{M}z}=0$, or $\beta_{\mathcal{M}|\tau}=0$. From
the mathematical point of view, the Rayleigh anomalies are the branch points; they set the conditions for the breaking of the incident angle or wavelength derivative of the transformation coefficients. In what follows we will mark the Rayleigh
anomalies as $(m_1,m_2)_\tau^R$ (which corresponds to the vanishing of $\beta_{\tau|m_1,m_2}$) or as $[m_1,m_2]_\tau^R$
(which corresponds to the vanishing of $\beta_{\tau|[m_1,m_2]}$).

Since modulation harmonics play a crucial role in SPP excitation, et us consider below the Fourier representation of
the structures predominantly used in the experiments.

\subsection{\label{subsec:harmonics}The modulation spectra} %Fourier harmonics

The majority of experimental works deal with hole arrays fabricated with the use of metal films deposited onto a
dielectric substrate (predominantly onto quartz), see Fig.~\ref{fig:geometry}.
Instead of holes we will consider a cylindrical inclusions in the film. The inclusions are supposed to consist of a metal or a semiconductor with the dielectric permittivity different from the permittivity of the film. Both the inclusions and the film must be highly conducting. This structure, as we will see below, may qualitatively describe the optical properties of nanohole arrays, even thought the film does not contain holes as such. Furthermore, the results from our studies may be used for modeling and describing arrays of cylindric nanoparticles, so far examined experimentally. Indeed, the structure considered may be fabricated by making an array of cylindric nanoparticles with impedance $\xi_i$, which are immerged into a conducting film.
\begin{figure}[!tb]
  \includegraphics[width=8cm]{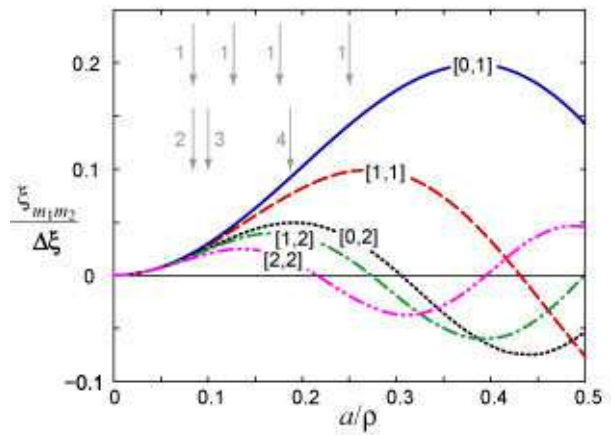}\\
  \caption{\label{fig:ksi(r0)}(Color online) Modulation harmonics $\xi_{m_1,m_2}$
dependence upon a ratio of the cylindrical inclusion radius to the period of the square array, $a/\rho$. The arrows 1,
2, 3, 4 indicate $a/\rho$ values of the hole arrays studied in Refs.~\onlinecite{Ebbesen98_Nature},
\onlinecite{Ebbesen_PRB_98}, \onlinecite{shadows_APL_02}, \onlinecite{Ebbesen_theory_PRL01} correspondingly.}
\end{figure}

For the inclusions of a round cross-section with radius $a$, we have
\begin{eqnarray}\label{ns1}
\tilde{\xi}_{\mathcal{M}}= \frac{2\pi a \Delta\xi}{\Sigma\cdot \mathrm{g}_{\mathcal{M}}}J_1(a\mathrm{g}_{\mathcal{M}}),
\q \xi_{\mathcal{O}}=\xi_f+\frac{\pi a^2 \Delta\xi}{\Sigma}
,\nonumber\\
\mathbf{g}_{\mathcal{M}}=m_1\mathbf{g}_{1}+m_2\mathbf{g}_{2}, \q \Sigma=|\boldit{\rho}_1\times \boldit{\rho}_2|,
\end{eqnarray}
where $\Delta\xi=\xi_f-\xi_i$ is the difference between  film ($\xi_f$), and inclusion ($\xi_i$) surface impedance,
$J_1$ is the first-order Bessel function. A typical example for the Fourier spectrum of the array of cylindrical
inclusions for the square symmetry structure is shown in Fig~\ref{fig:ksi(r0)}. The square symmetry structure means,
that the angle between translation vectors is $\pi/2$ and their modules are equal, $\rho_1=\rho_2$. In the following
calculations we take $\xi_f$ to be equal  to the impedance of $Ag$ (using the wavelength dependence from
Ref.~\onlinecite{Book_Zolotarev84}); the impedance of inclusions will be modeled as $\xi_i=\mathrm{w}\xi_f$, where
$\mathrm{w}$ is the dimensionless parameter.

As evident from \equ{ss1}, the amplitude of the resonance ($\mathcal{R}$th) diffraction order is proportional to
$\tilde{\xi}_{\mathcal{R}}$, $H^{\tau|\sigma}_{\mathcal{R}}\propto\tilde{\xi}_{\mathcal{R}}$, therefore, the efficiency
of excitation of SPP in this order is strongly dependent upon the corresponding ``resonance'' amplitude
$\tilde{\xi}_{\mathcal{R}}$. For large values of $r_1,r_2$ the Fourier amplitude $\tilde{\xi}_{\mathcal{R}}$ tends to
zero as
\begin{equation}\label{ns1.1}
  \begin{split}
\tilde{\xi}_{\mathcal{R}}\sim \mathrm{g}_{\mathcal{R}}^{-3/2}, \q |\mathcal{R}|\gg1.
\end{split}
\end{equation}
Thus, the resonances in high diffraction orders are less efficient. If the inclusion
diameter is far smaller than the modulation periods, the low-order amplitudes are
approximately independent of their order:
\begin{equation}\label{ns2}
  \begin{split}
\tilde{\xi}_{\mathcal{M}}\simeq \pi \Delta\xi a^2\Sigma^{-1}[1+O(a\mathrm{g}_{\mathcal{M}})] \q \mathrm{for} \q
a\mathrm{g}_{\mathcal{M}}\ll1.
\end{split}
\end{equation}
So, for very thin inclusions the Fourier amplitudes of the structure show a slight decrease, which is valid for an
arbitrary cross-section of the inclusions. In its turn, it significantly broadens the SPP resonances:
they become ``diffusive'' due to different scattering processes of the excited SPPs.

Let us write also the Fourier expansion of the structure having the inclusions of a rectangular cross-section,
\begin{equation}\label{rh1}
  \begin{split}
\tilde{\xi}_{\mathcal{M}}=\frac{4a_1a_2\Delta\xi}{
(\mathbf{a}_1\cdot\mathbf{g}_{\mathcal{M}})(\mathbf{a}_2\cdot\mathbf{g}_{\mathcal{M}})\Sigma}
\sin\frac{\mathbf{a}_1\cdot\mathbf{g}_{\mathcal{M}}}{2} \sin\frac{\mathbf{a}_2\cdot\mathbf{g}_{\mathcal{M}}}{2},
\end{split}
\end{equation}
where $\mathbf{a}_{1,2}$ are the vectors oriented along the sides of the rectangular
inclusions with modulus equal to their lengthes, $a_{1,2}$.

\section{\label{sec:nonsymmetric}SB SPP resonances. Nonsymmetric surrounding}

\begin{figure}[!tb]
  \includegraphics[width=8cm]{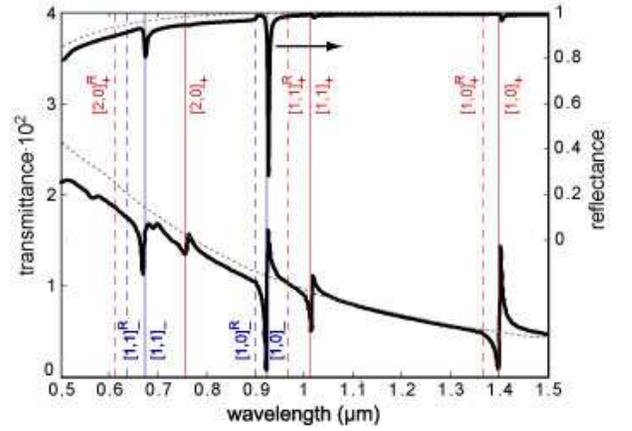}\\
  \caption{\label{All_NS}(Color online) Dependence of transmittance and reflectance
upon the wavelength for the wave normally-incident from the air superstrate onto the silver film bounding with the
quartz substrate. The parameters of $C_{4v}$ array, are $\rho_1=\rho_2=\rho=0.9\mu m$, $a/\rho=1/3$. The incident wave
electric field $\mathbf{E}$ is parallel to $\mathrm{\mathbf{g}}_1$ (to $Ox$ axis, see. Fig.~\ref{fig:geometry}). The
inclusion impedance, $\xi_i$, is taken to be $\xi_i=2\xi_f$. The film thickness is equal to $2.3$ skin depths.
The transmittance and reflectance of the unmodulated film are plotted with dotted lines. The vertical solid and dashed lines indicate the wavelengths relevant to SPPs on the unmodulated boundaries and Rayleigh points respectively.}
\end{figure}

Bearing in mind the experiments already made, periodically-modulated films surrounded by two different dielectric media
are of special interest. These structures can support both SB-localized SPPs and, under the special conditions,
DB-localized SPPs. The latter will be discussed separately. In the vicinity of normal incidence we mainly focus on, the resonances basically correspond to SB-localized SPPs. An example of the transmittance and
reflectance wavelength spectra for the strictly normal incidence is illustrated in Fig.~\ref{All_NS}. The calculations
are for $C_{4v}$ array of inclusions with a round cross-section in the Ag film surrounded by air
superstrate and quartz substrate. The transmittance and reflectance extremes are pronounced at wavelengths
for the SPPs existing on unmodulated metal-quarts and metal-air interfaces. Note that while the extremes are
shifted with respect to initial SPPs wavelength, wich is due to the modulation and finite film thickness influence,
Rayleigh anomalies are unshifted.

\begin{figure*}[!t]
  \includegraphics[width=17cm]{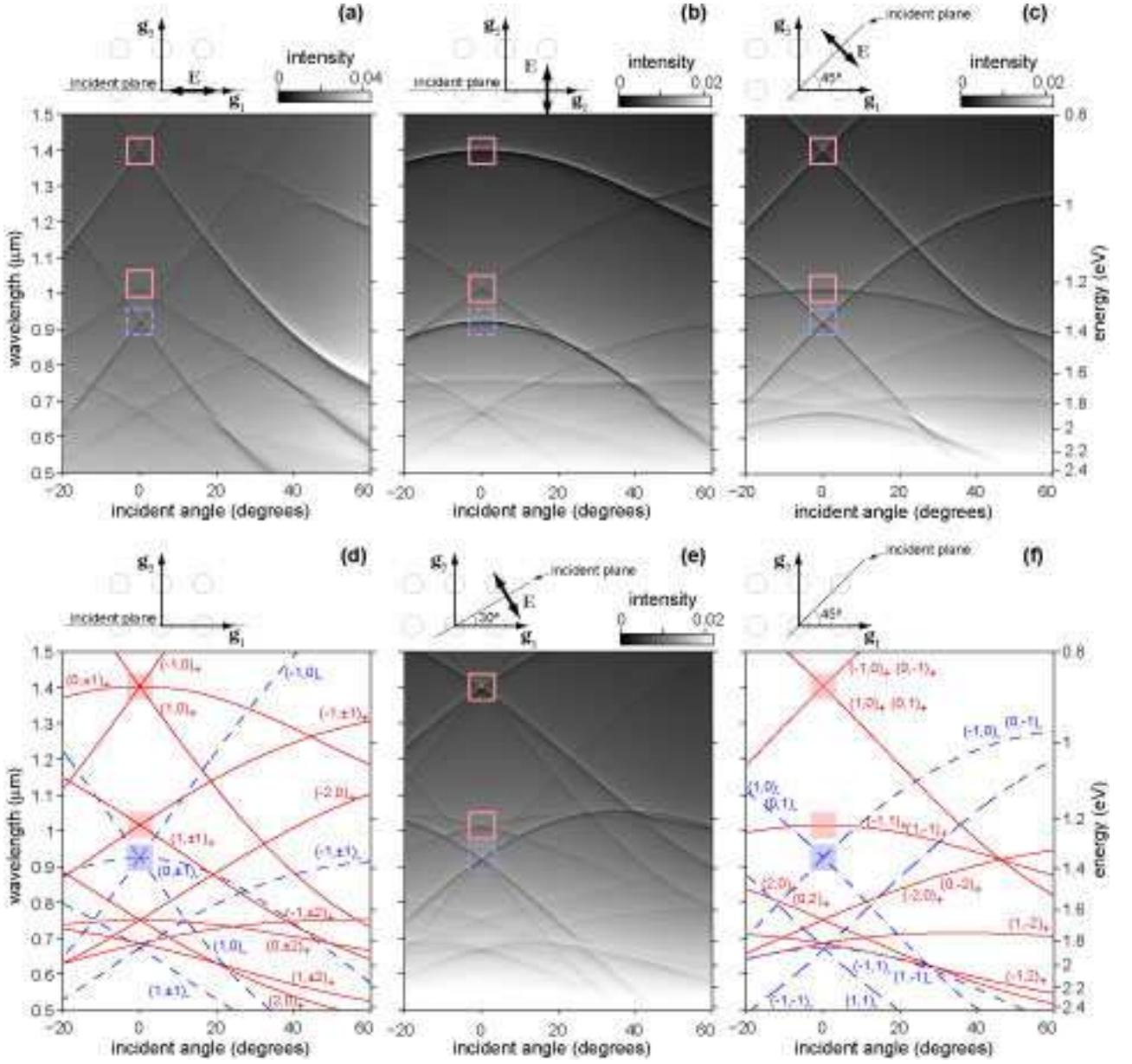}\\
  \caption{\label{t2NS}(Color online)
Dependence of transmittance upon the wavelength (energy) and the incident angle for different polarization of the
incident wave and different orientations of the incident plane. The array and the film parameters are the same as in
Fig.~\ref{All_NS}. In (d, f) solid and dashed lines are the resonance curves, Eq.~\eqref{45.2.0}. Solid (dashed)
squares indicate the vicinity of the quartz-silver (air-silver) resonances which are discussed in detail in the text.}
\end{figure*}

Now discuss the vicinity of the strictly normal incidence. In Figs.~\ref{t2NS} (a-c) and (e) the transmittance
contour plot is shown as a function of the wavelength or/and photon energy and the incident angle for different
polarization of the incident wave. In Fig.~\ref{t2NS}  (d,f) the resonance curves \eqref{45.2.0} corresponding to
the contour plots in Fig.~\ref{t2NS}  (a-c) are presented.
There is an excellent agreement between the transmittance features and the resonance curves. It should be stressed that
the short-wavelength region  corresponds to strong dispersion, $\xi(\lambda)$. In Figs.~\ref{t2NS} (a-c) the special
symmetry instances are shown. Viz., orientation of the incident plane relative to one of the reciprocal grating vectors
(in our case it is $\mathbf{g}_1$) at angles $0$ (or $\pi/2$) and $\pi/4$ results in the coincidence of the resonance
curves, see discussion below formula \eqref{45.2.0}. The coincidence is appreciable in the vicinity of $[1,0]_+$,
$[1,1]_+$ and $[1,0]_-$ resonances. In Fig.~\ref{t2NS} these vicinities are marked by squares. We see that in
Figs.~\ref{t2NS} (a-c) from one to three resonance features belong to these regions. Evidently, in Figs.~\ref{t2NS}
(d,f) from one to three resonance curves resonance intersect these regions. Alternatively, four resonance ``mountain
ridges'' intersect in the vicinity of a single point, see Fig.~\ref{t2NS} (e), which displays the transmittance contour
plot for the nonspecific in the angle $\psi$ case.

Emphasize, that the intersections of solid and dashed resonance curves in Figs.~\ref{t2NS} (d,f) and the intersections
of corresponding resonance ``mountain ridges'' in the transmittance contour plots are due to DB SPP resonances. The
latter comply with the excitation of dressed SPPs, resulting from the coupling of initial SPPs of different film faces
through the modulation.

In the following subsections we examine in detail some of SB fourfold resonances.

\subsection{\label{subsec:[10]+N}$[1,0]_+$ resonance}

Consider first the longest-wavelength resonance (at 1.4 $\mu m$ in Fig.~\ref{All_NS}) in the vicinity of
close-to-normal incidence. This resonance arises due to excitation of the fourfold $[1,0]_+$ SB SPP at the metal-quartz
interface having $C_{4v}$ modulation symmetry.

\begin{figure*}[!t]
  \includegraphics[width=17cm]{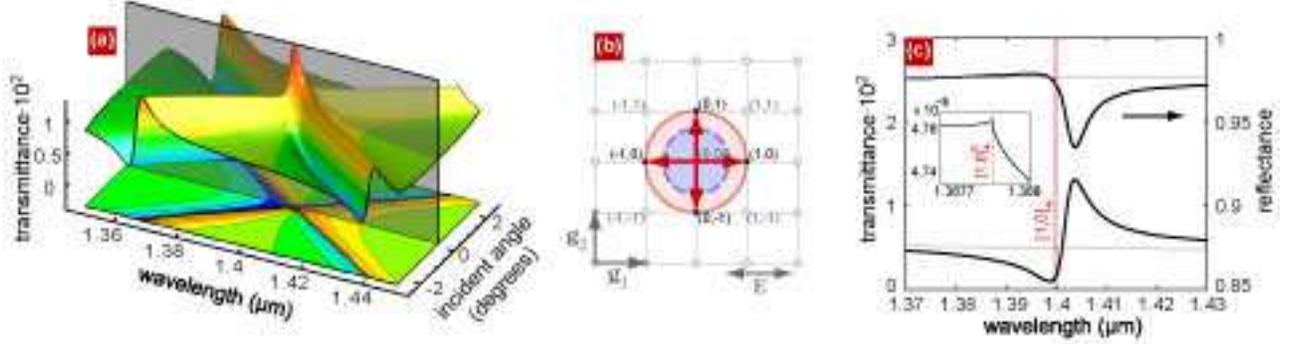}\\
  \caption{\label{[10]+}(Color online) (a) The
dependence of the zeroth-order transmittance, $\tau_{\mathcal{O}}$, upon the
wavelength and incident angle in the vicinity of $[1,0]_+$ resonance on silver
array of inclusions with the symmetry $C_{4v}$ for  $\mathbf{E} \|
\mathbf{g}_1$. In (b) the reciprocal grating is shown, where the radii of solid
and dashed circles are equal to the SPP wave vector modulus in quartz and air
half-spaces, respectively; the circles separate the regions of propagating (filled
area) and evanescent waves in the superstrate and substrate. In (c) the
wavelength dependence for the zeroth-order transmittance, $\tau_{\mathcal{O}}$,
and reflectance, $\rho_{\mathcal{O}}$, are shown, which correspond to the grey
section in (a). The inset shows the vicinity of the Rayleigh anomaly.
The parameters of the array are the same as in Fig.~\ref{All_NS}.}
\end{figure*}

In Fig.~\ref{[10]+} (a, c) the enlarged fragments of Figs.~\ref{All_NS} and \ref{t2NS} (a) are shown in  the vicinity
of $[1,0]_+$ resonance for the incident plane orientated parallel to $\mathbf{g_1}$ and the polarization of the
incident wave such that $\mathbf{E} \| \mathbf{g}_1$. Note that the resonance features are heavily dependent upon the
polarization of the incident wave and the orientation of the incident plane, cf. transmittance dependencies shown in
Fig.~\ref{t2NS} (a), (b), (c) and (e). For instance, for $p$ polarization, when $\mathbf{E}$ belongs to the incident
plane parallel to $\mathbf{g}_1$ ($\psi=0$), in the vicinity of normal incidence, only the pair of the resonance waves
are excited, see Subsection~\ref{subsec:multiple}. These waves correspond to $(\pm1,0)$ diffraction orders (TCs
$T^{+|--}_{0,\pm1}$ are equal to zero).  In Fig.~\ref{t2NS}~(a) and in Fig.~\ref{[10]+}~(a) one can observe the
intersection of the two resonance features, relevant to the intersection of two resonance curves: $(1,0)_+$ and
$(-1,0)_+$, Fig.~\ref{t2NS}~(d). In deflecting from the normal incidence, the projection of $\mathbf{H}_t$ onto
$\mathbf{H}^+_{0,\pm1}$ becomes nonzero, and it appears as an feature in Fig.~\ref{t2NS} (a) close to the coinciding
resonance curves $(0,\pm1)_+$ in Fig.~\ref{t2NS} (d). When $\mathbf{E}$ is perpendicular to the incident plane parallel
to $\mathbf{g}_1$ ($\psi=0$), [$s$-polarization, see Fig.~\ref{t2NS}~(b)], the only excited pair of resonance waves
comply with the diffraction orders $(0,\pm1)$ (resonance coefficients $T^{+|--}_{0,\pm1}$). If the incident plane is
oriented at $\psi\neq0,\pi/4$, then   two pairs of resonance waves are excited both for $p$- and $s$-polarizations of
the incident wave in  the vicinity of normal incidence [in Fig.~\ref{t2NS}~(c,e) $s$-polarization case is shown]. The
resonance waves correspond to diffraction orders $(\pm1,0)$ and $(0,\pm1)$. In Fig.~\ref{t2NS}~(f) this appears as the
intersection of four resonance curves: $(\pm1,0)_+$ and $(0,\pm1)_+$, coinciding for $\psi=\pi/4$.

The wavelength dependence of the zeroth-order wave has the typical Fano-type profile which consists of the neighbouring
minima and maxima, see Fig.~\ref{[10]+}~(c). This is due to the interference (see Subsection~\ref{subsec:single})
between nonresonance and resonance transmittance mechanisms. The maxima of the Fano profile is red-shifted as compared
with the wavelength of the SPP existing at the nonmodulated metal-quartz interface. This shift is mainly due to the
scattering by modulation through the diffracted inhomogeneous waves, and partially due to the finite film thickness.

\begin{figure}[!hbt]
  \includegraphics[width=8cm]{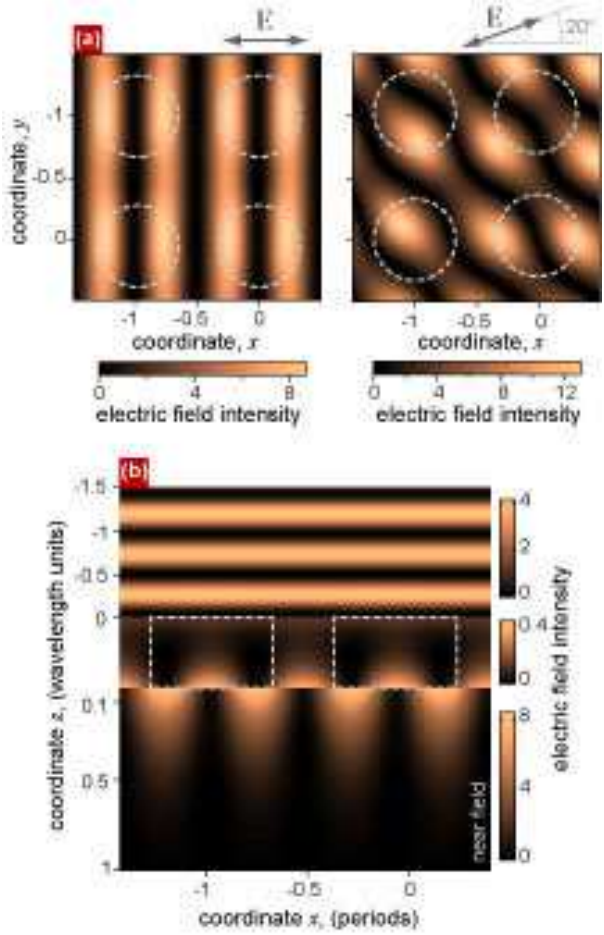}\\
  \caption{\label{[10]+field}(Color online) The
squared field distribution is shown for $[1,0]_+$ SPP resonance at $\lambda=1.4023~\mu m$, $\theta=0$. The inclusions
are shown by the dashed lines. The parameters of the array are the same as in Fig.~\ref{All_NS}.}
\end{figure}

The amplitude of the excited resonance waves is proportional to $\tilde{\xi}_{[10]}$ Fourier harmonic amplitude, while
the zeroth-order transmittance and reflectance are proportional to its squared value, as it follows from
Eqs.~\eqref{ss1}-\eqref{ss7} (see also Appendix~\ref{C2v}). Therefore, this harmonic plays the most important role in
the excitation of the SPP eigenmode for $[1,0]_+$ resonance. Actually, the structure of the eigenmodes in the modulated
film (dressed SPPs) define  the distribution of the near-field completely. As indicated in
Subsection~\ref{subsec:multiple}, the only excited eigenmodes are high-frequency ones. For instance, when $\mathbf{E}$
is oriented along $\mathbf{g}_1$, the distribution of the squared  field on the quartz side take the form
$|\mathbf{E}^+|^2\sim \sin^2[\boldrm{k}_{(1,0)t}\mathbf{r}]=\sin^2(\mathbf{g}_{1}\mathbf{r})$, similary to the spatial
field distribution for the twofold coupling given by Eq.~\eqref{mr2}. Indeed, this is in compliance with the left-hand
part of Fig.~\ref{[10]+field} (a). In the general case, when $\mathbf{E}$ is oriented relative to $\mathbf{g}_1$ at an
arbitrary angle $\psi$, the near-field structure is formed by the interference of the eigenmodes. One of them
corresponds to $\sin(\mathbf{g}_{1}\mathbf{r})$ and the other is for $\sin(\mathbf{g}_{2}\mathbf{r})$. As a result the
field on the interface has the form $|\mathbf{E}^+|^2\sim
[\chi_1\sin(\mathbf{g}_{1}\mathbf{r})+\chi_2\sin(\mathbf{g}_{2}\mathbf{r})]^2$, where the ``weight'' coefficients
$\chi_1$ ($\chi_2$) are proportional to $\cos\psi$ ($\sin\psi$) respectively. This interference is clearly seen in the
right-hand part of Fig.~\ref{[10]+field} (a). Thus the near-field structure is in excellent agreement with the
experiment, cf. Ref.~\onlinecite{shadows_APL_02}.

Consider the field distribution along $z$ axis, see Fig.~\ref{[10]+field} (b).
Since the far-side SB SPP resonances provide small reflectance dips [the lowest reflectance is of order of 93\%, see
Fig.~\ref{[10]+} (c)], we notice a clear interference pattern along $z$ axis in the air half-space. The field
penetrating into the film decays exponentially, and excites a SB SPP at the quartz-metal interface. The SPP field has a
well-pronounced ``torch'' structure. It decays exponentially into the film at a distance of order of skin-depth,
$\delta$, where $\delta/\lambda\sim1/|\sqrt{\overline{\varepsilon}}|\ll1$ and decays into the dielectric medium
(quartz) at a distance of $\delta_+$, where
$\delta_\tau/\lambda\simeq\sqrt{|\overline{\varepsilon}/\varepsilon_\tau|}/2\pi\sim1$. Note that the field pattern
along $x$ axis in the film is shifted to a half-period with respect to the pattern in the quartz. This is because the
SPP electric field in the metal has predominantly tangential-to-interface component as compared with a dielectric where
the electric field has predominantly normal to the interface component. The intensity of the far field in the quartz
half-space is constant, since there exists only one propagating wave (at zeroth diffraction order).

\subsection{\label{subsec:polariz[10]+N}Polarization properties of $[1,0]_+$ resonance:
dependence upon the inclusion shape}

In this subsection we give a  theoretical interpretation of the recent
experiments\cite{hole_shape_PRL04,shape_localized_PRB05} concerning the influence of the hole shape on ELT through
periodic hole arrays within the framework of our theory. Moreover, we describe the transmittance behavior while
changing the polarization of the incident light.

Consider a  silver film sandwiched between an air superstrate and quartz substrate with $C_{2v}$ modulation symmetry
caused by the inclusions of rectangular cross-section. This symmetry reduces to $C_{4v}$ symmetry for a circle
cross-section of inclusions. Now let us give a close consideration to the $[1,0]_+$ SB fourfold SPP resonance at
strictly normal incidence. As was discussed above, for $C_{2v}$ symmetry structures the polarization of the transmitted
radiation depends upon the polarization of the incident wave. Suppose that the direction of $\mathbf{g}_1$ is parallel
to the small side of the rectangular, $a_1$. If  the incident-wave electric field is oriented along $\mathbf{g}_1$
[$\mathbf{g}_2$], the incident wave magnetic field is then matched with that of $(\pm1,0)_+$ [$(0,\pm1)_+$] SPP pairs
having wavevectors $\pm\mathbf{g}_1$ [$\pm\mathbf{g}_2$] for normal incidence, see Fig.~\ref{[10]+} (b). According to
our approach, the transmittance coefficient, $\tau_\mathcal{O}$, is proportional to $\tilde{\xi}_{\pm1,0}^4$
($\tilde{\xi}_{0,\pm1}^4$) with a of polarization along $\mathbf{g}_1$ ($\mathbf{g}_2$). This leads to
$\tau_\mathcal{O}\sim(\Delta\xi)^4(a/\rho)^4J_1^4(2\pi a/\rho)$ for round cross-section and
$\tau_\mathcal{O}\sim(\Delta\xi)^4(a_2/\rho)^4\sin^4(\pi a_1/\rho)$
[$\tau_\mathcal{O}\sim(\Delta\xi)^4(a_1/\rho)^4\sin^4(\pi a_2/\rho)$] for rectangular cross-section with $\mathbf{E}$
orientation along $\mathbf{g}_1$ ($\mathbf{g}_2$) [see Eqs.~\eqref{ns1},~\eqref{rh1}]. We take the same geometrical
parameters for periodic arrays as in Ref.~\onlinecite{hole_shape_PRL04}, i.e. $\rho=425~nm$; $a_1\times a_2=
75\times225~nm^2$ for small rectangles and $a_1\times a_2= 150\times225~nm^2$ for large ones; and $2a=190~nm$ for the
circle diameter which corresponds to the resonance position at $\lambda\simeq0.72~\mu m$. Taking the film thickness to
be equal to 3 skin depths, $\Phi'=3$, and the inclusion impedance to be $\xi_i=2\xi_f$, where $\xi_f$ is taken for the
silver, we obtain the transmittance of order $2\%$, which is nearly 7 times higher than the transmittance through the
unmodulated film.

\begin{figure}[!htb]
  \includegraphics[width=8cm]{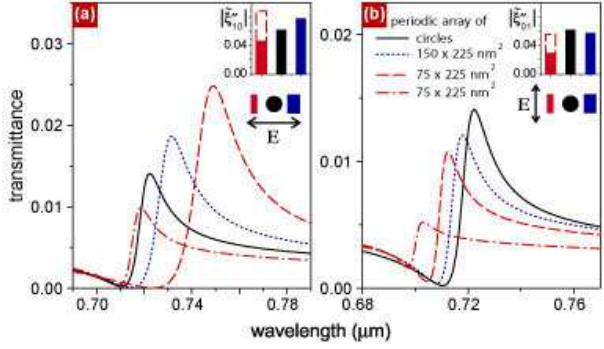}\\
  \caption{\label{polar[10]+} (Color
online) The wavelength dependence of transmittance for the inclusion
with different cross-sections at strictly normal incidence onto $Ag$
film deposited onto the quartz substrate. The polarization of the
incoming light, $\mathbf{E}$, is perpendicular to the long axis of
the rectangles in (a) and parallel to the long axis of the
rectangles in (b). The dotted curves are for the large rectangular
inclusions. The dashed and dash-dotted curves are for the small
rectangular inclusions. The solid curves correspond to the cylindrical
inclusions. The film thickness is equal to 3 skin depths. In the
insets the amplitude of the $(\pm1,0)$ and $(0,\pm1)$ modulation
harmonics are shown. The dashed rectangulars in the insets indicate
the modulation amplitude for the dashed curves. }
\end{figure}

When $\mathbf{E}$ is directed along $\mathbf{g}_1$, the zeroth-order transmittance amplitude is proportional to a
squared $\tilde{\xi}_{\pm1,0}$ harmonic amplitude. This amplitude increase for large rectangles as compared with the
circles and this results in redshift and in a rise of the transmittance maximum, see Fig.~\ref{polar[10]+}~(a).
This tendency is in accordance with the experiment.  But for the small rectangles the resonance harmonic amplitude is
the smallest one and this leads to blueshift and a decrease in the transmittance maxima as compared with the circles
and large rectangles. This is not what was observed experimentally. We may eliminate this discrepancy only by adjusting
the impedance of the inclusion. We use this adjustment because, in our approximation, we do not allow for the influence
of the modulation upon the volume of the film, but the modulation appears in the boundary conditions only. In other
words, the fields decay into the film with the same decrement at any cross-section by a plane parallel to axis $Oz$. On
the other hand, it is evident that under a strong modulation, with the impedances of inclusions and the film differing
greatly, the fields in the inclusion and in the volume of the film decay in a different way. For instance, inside the
holes the field show a weaker decay than in the metal regions. Thus, adjusting the impedance, we have found that for
$\xi_i>2.67\xi_f$ the transmittance maximum corresponding to the small rectangles is redshifted and increased as
compared with the large rectangulars and circles as it was in the experiment. If we rotate the polarization plane by
$\pi/2$ so that $\mathbf{E}$ is directed along $\mathbf{g}_2$, see Fig.~\ref{[10]+} (b), the transmittance will depend
upon the inclusion shape differently. In Fig.~\ref{polar[10]+}~(b) the transmittance is shown for different inclusion
shapes both with equal $\xi_i$ for all shapes and when the modulation amplitude is taken from the previous case for the
small rectangles.

On the one hand, supposing the inclusion impedance, $\xi_i$, to be constant, but at the same time
changing the inclusion shape, we can predict the true transmittance through arrays of nanoparticles
immersed into the film (each nanoparticle corresponds to an inclusion).
On the other hand, by adjusting the inclusion impedance for different shapes, we may
provide the qualitative coincidence with experimental measurements of ELT through hole
arrays.

Consider the experiment of Ref.~\onlinecite{shape_localized_PRB05}, where the effect of aspect ratio of rectangular
holes on the transmittance of periodic arrays of subwavelength holes in optically thick metal films was measured. It
was found that as the electric field, $\mathbf{E}$, of the incident light is directed along the short axes of the
holes, $a_1$, the zeroth-order transmittance peak (corresponding to $[1,0]_+$ SPP resonance) increases and suffers a
redshift when the aspect ratio of the holes, $a_2/a_1$, is enlarged. Conversely, it was discovered that as
$\mathbf{E}$ is directed along the long axes of the holes, $a_2$, $[1,0]_+$ transmittance peak decreases and undergoes a blueshift when the aspect ratio of the holes is enlarged. The experimental measurements was also confirmed by strict
numeric calculations. The authors attribute the discovered polarization dependence of the transmittance upon the the
ratio of the rectangular holes to a result from interaction between SPP resonances at the surface of the metal and shape resonances inside the holes.

We hold to a different, simpler point of view. It is obvious that the localized eigenmodes are not supported by the
structures under examination. But the calculations we do not present here within the framework of our simple model
exhibit the same behavior of the transmittance, as in Ref.~\onlinecite{shape_localized_PRB05}. We have taken the
geometrical parameters of the inclusions the same as in Ref.~\onlinecite{shape_localized_PRB05}. When $\mathbf{E}$ is
parallel to $a_1$, the amplitude of the excited $(\pm1,0)_+$ SPP is proportional to $\tilde{\xi}_{\pm1,0}$, which
increases as the aspect ratio increases. Therefore this leads to the transmittance peak growth and redshift. On the
contrary, when $\mathbf{E}$ is parallel to $a_2$, the amplitude of the excited $(0,\pm1)_+$ SPP is proportional to
$\tilde{\xi}_{0,\pm1}$, which decreases with an aspect ratio increase. This brings about the transmittance peak
decrease and blueshift. Thus, the polarization behavior may be successfully described under the assumption of the
excitation of purely interface SPPs.

\subsection{\label{subsec:[11]+N}$[1,1]_+$ resonance}

\begin{figure*}[!t]
  \includegraphics[width=17cm]{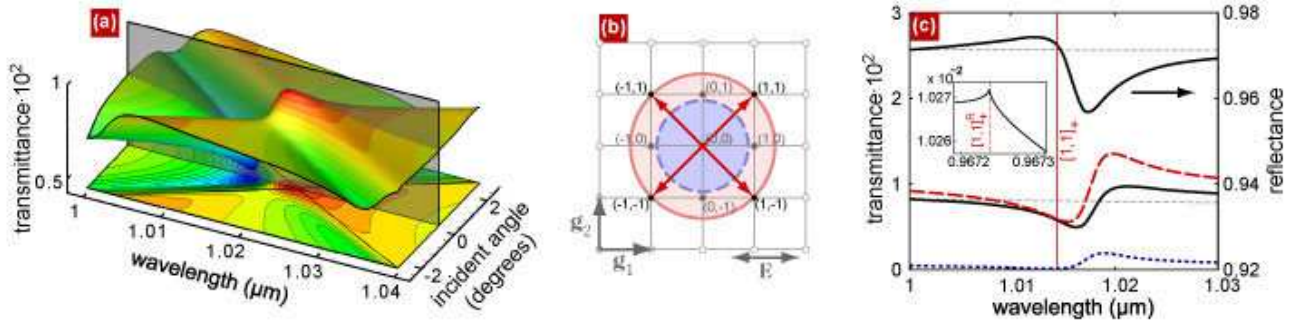}\\
\caption{\label{[11]+}(Color online) (a) The dependence of the zeroth-order  transmittance, $\tau_{\mathcal{O}}$, upon
the wavelength and incident angle corresponding to $[1,1]_+$ resonance on $C_{4v}$ silver array of cylindrical
inclusions, $\mathbf{E} \| \mathbf{g}_1$. In (b) the reciprocal grating is shown [compare with Fig.~\ref{[10]+} (b)].
In (c) the wavelength dependence for the zeroth-order transmittance, $\tau_{\mathcal{O}}$, and reflectance,
$\rho_{\mathcal{O}}$, (solid curves) along with the nonzeroth-order transmittance, $\tau_{(\pm1,0)}$, (dotted curve)
and total transmittance, $\tau_{\mathrm{tot}}$, (dashed curve) are shown. It corresponds to section of (a) along the
plane, marked in grey. The inset presents the Rayleigh anomaly. The parameters of the array are the same as in
Fig.~\ref{All_NS}.}
\end{figure*}

Now examine the resonance near $1~\mu m$ ($1.65~eV$) in the vicinity of close-to-normal incidence, see
Figs.~\ref{All_NS}, \ref{t2NS}. This fourfold $[1,1]_+$ resonance is consistent with excitation of a SB SPP at the
metal-quartz interface, see Fig.~\ref{[11]+} (b). $\lambda-\theta$ dependence is shown in Fig.~\ref{t2NS} (square
regions) for different polarizations and orientations of the incident plane. Fig.~\ref{[11]+} (a) and (c) are the
enlarged fragments of Fig.~\ref{t2NS} (a) and Fig.~\ref{All_NS}.

\begin{figure}[!hbt]
  \includegraphics[width=7.5cm]{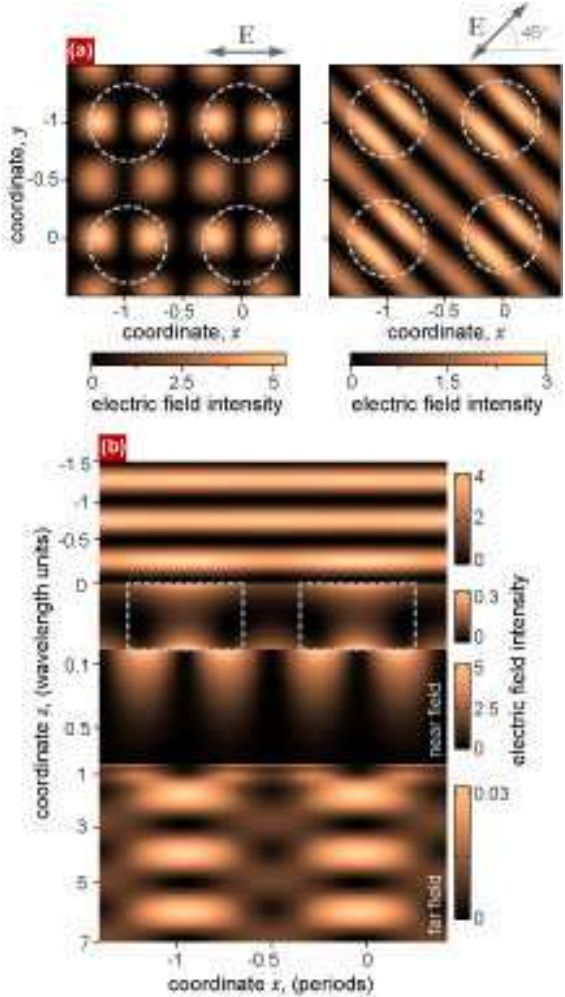}\\
  \caption{\label{[11]+field}
The squared field distribution for $[1,1]_+$ SPP resonance is shown
at $\lambda=1.0168 \mu m$, $\theta=0$. The parameters of the array
are the same as in Fig.~\ref{All_NS}}
\end{figure}

Now we explain the dependence of the transmittance upon $\lambda$ and $\theta$ in the vicinity of $[1,1]_+$ resonance,
see the second (from above) solid square regions in Fig.~\ref{t2NS}. For $p$ polarization, when $\mathbf{E}$ belongs to
the incident plane parallel to $\mathbf{g}_1$ ($\psi=0$),  both the pairs of the resonance waves corresponding to
$(\pm1,\pm1)$ and $(\mp1,\pm1)$ diffraction orders are excited in the close-to-normal incidence region. Accordingly, in
Fig.~\ref{t2NS}~(a) and in Fig.~\ref{[11]+}~(a) one can see the intersection of the two features that are relevant to
two pairs of coinciding resonance curves, $(1,\pm1)_+$ and $(-1,\pm1)_+$, in Fig.~\ref{t2NS}~(d). Evidently, the
similar dependencies take place if $\mathbf{E}$ is perpendicular to the incident plane parallel to $\mathbf{g}_1$
($\psi=0$) [$s$-polarization, see Fig.~\ref{t2NS}~(b)]. If the incident plane is oriented at $\psi=\pi/4$, then for
$s$-polarization of the incident wave [see Fig.~\ref{t2NS}~(c)] the magnetic field $\mathbf{H}$ is parallel to
$\mathbf{g}_2-\mathbf{g}_1$. And the only excited pair of resonance waves in the vicinity of normal incidence
corresponds to the diffraction orders $(\pm1,\mp1)$ (resonance coefficients $T^{+|--}_{\pm1,\mp1}$). This pair provides
an extremum of the transmittance close to the coinciding resonance curves $(\pm1,\mp1)_+$ in Fig.~\ref{t2NS}~(f). For
the nondegenerated case, in the terms of classifying the resonance curves, Fig.~\ref{t2NS}~(e), there is the
intersection of four features in the transmittance which conforms to four distinct resonance curves: $[1,1]_+$.

The principal difference of $[1,1]_+$ resonance from the $[1,0]_+$ resonance considered above is that additional
propagating waves exist in the quartz substrate. Namely, the waves having $[1,0]$ diffraction orders provide the
nonzeroth-order ELT\cite{Ours_JETPHL_04,film_PhysRev} due to the scattering of the resonance waves through amplitudes
$\tilde{\xi}_{[1,0]}$. This scattering contributes additionally to the resonance width. Black dots, placed within the
solid circle in Fig.~\ref{[11]+}~(b) indicate the positions for tangential components of wavevectors of these waves. On
the other hand, the resonance $[1,1]_+$ is less pronounced as compared with $[1,0]_+$ one not only because of
additional radiation losses. As one can see from Fig.~\ref{fig:ksi(r0)}, the amplitude $\tilde{\xi}_{[1,1]}$, which is
responsible for resonance waves excitation within the limits of the main approximation, is smaller than the amplitude
$\tilde{\xi}_{[1,0]}$, and hence the SPP excitation is less effective. It should be noted that the SPP is excited not
only owing to first-order scattering of the incident wave by $[1,1]$ resonance harmonics, but also due to the
higher-order scattering processes. While the cubic processes (proportional to $\tilde{\xi}^3$) can be neglected, the
second-order scattering may compete with the linear one in some regions of parameter $a/\rho$, going beyond the
approximation \eqref{r}-\eqref{r1}. Thus, for $a/\rho>0.35$, the contribution to the amplitude of the resonance wave
from the second order term, proportional to $\tilde{\xi}_{[1,0]}^2$, is of the the same order as the contribution from
the linear term proportional to $\tilde{\xi}_{[1,1]}$ .

All the propagating field harmonics contribute to the transmitted energy flux in the substrate media. While the parts of the energy flux relating to the zeroth and nonzeroth diffraction orders are well separated at a grate distance from the film, they are overlapped in the region sufficiently close to the film. Therefore, when making the
experimental measurements, the detector position is crucial: if it is far away from the film, it fixes either
zeroth-order flux, $\tau_{\mathcal{O}}$ [see Fig.~\ref{[11]+}~(c), solid curve], or nonzeroth one, $\tau_{[1,0]}$ [see
Fig.~\ref{[11]+}~(c), dotted curve], depending upon detector orientation relative to the film. In case the detector is in the close-to-the-film vicinity, it fixes the total flux [see Fig.~\ref{[11]+}~(c), dashed curve]. The interference pattern resulting from the existence of several homogeneous waves is quite noticeable in the far-field region in
Fig~\ref{[11]+field}~(b). The near-field structure may be understood in the same way as in the case of $[r,0]$ resonance. For instance, when $\mathbf{E}$ is oriented along $\mathbf{g}_1+\mathbf{g}_2$, the intensity distribution on the quartz side for strictly normal incidence is proportional to $|\mathbf{E}^+|^2\sim
\sin^2[\boldrm{k}_{(1,1)t}\mathbf{r}]=\sin^2[(\mathbf{g}_1+\mathbf{g}_2)\mathbf{r}]$, see the right-hand part of
Fig.~\ref{[11]+field} (a). When $\mathbf{E}$ is oriented with respect to $\mathbf{g}_1+\mathbf{g}_2$ at $\psi=\pi/4$, the near-field structure is formed by the interference of the eigenmode for $\sin[(\mathbf{g}_1+\mathbf{g}_2)\mathbf{r}]$ and the eigenmode for $\sin[(\mathbf{g}_1-\mathbf{g}_2)\mathbf{r}]$, so that the field at the interface takes the form $|\mathbf{E}^+|^2\sim
\{\sin[(\mathbf{g}_1+\mathbf{g}_2)\mathbf{r}]+\sin[(\mathbf{g}_1-\mathbf{g}_2)\mathbf{r}]\}^2$. This is in accordance
with the left-hand part of Fig.~\ref{[11]+field} (a). It is interesting to note that the field intensity is increased
additionally in the regions where the inclusions are located.

\subsection{\label{subsec:[10]-N}$[1,0]_-$ resonance}

It should be emphasized, that SB SPP resonances pertinent to the  superstrate face provide strong nonzeroth-order ELT
on condition that $\varepsilon_-<\varepsilon_+$. It is clear from the geometrical reasons. Since the modulus of the
wavevectors in the superstrate media are less than those of the substrate media, the metal-superstrate SPP wavevector
($k_{SPP}^-\simeq k\sqrt{\varepsilon_-}$) corresponds to the propagating wave in the substrate media. This enables SPP
leakage without scattering via the modulation. Another feature of superstrate resonances is far deeper minima in the
zeroth-order reflectance as compared to the resonance on the substrate face. This is because the zeroth-order
reflectance is strongly dependent upon the efficiency of SPP excitation. The incident light excites SPP on the
superstrate face directly via the modulation, while for excitation on the substrate face, amplitude of the light
decreases prior to the excitation caused by the tunneling through the film (thereby making the excitation process less
effective).

\begin{figure*}[!t]
  \includegraphics[width=17cm]{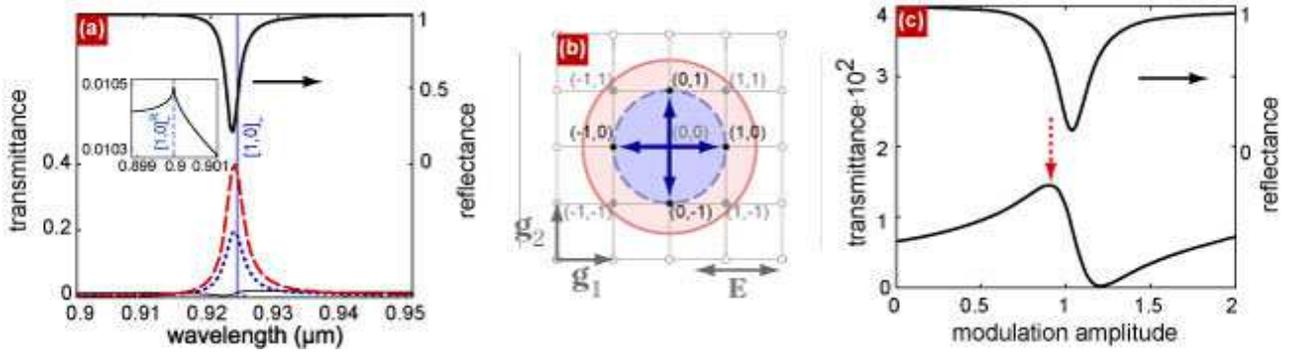}\\
  \caption{\label{[10]-}(Color online) In (a) the wavelength dependence for the
zeroth-order transmittance, $\tau_{\mathcal{O}}$, and reflectance, $\rho_{\mathcal{O}}$, (solid curves) along with the
nonzeroth-order transmittance, $\tau_{(\pm1,0)}$, (dotted curve) and total transmittance, $\tau_{\mathrm{tot}}$,
(dashed curve) are shown for $[1,0]_-$ resonance on silver $C_{4v}$ array of inclusions, $\mathbf{E} \| \mathbf{g}_1$,
$\theta=0$. In the inset the Rayleigh anomaly is shown. (b) The reciprocal grating [compare with Figs.~\ref{[10]+}
(b),~\ref{[11]+} (b)]. (c) The dependence of $\tau_{\mathcal{O}}$, and $\rho_{\mathcal{O}}$ upon the modulation
amplitude, $\mathrm{w}=|(\xi_f-\xi_i)/\xi_f|$. The parameters of the array are the same as in Fig.~\ref{All_NS}.}
\end{figure*}

Let us discuss the SB metal-air $[1,0]_-$ SPP resonance that corresponds to the transmittance (reflectance) maximum
(minimum) at $\lambda\simeq0.92 \mu m$ in Fig.~\ref{All_NS}. The enlarged fragment of the wavelength-dependent zeroth-
and nonzeroth-order transmittances are shown in Fig.~\ref{[10]-}~(a). While the zeroth-order transmittance profile is
similar to that corresponding to the $[1,0]_+$ resonance [their maximal amplitudes are practically equal, being of
order $\tau_{\mathcal{O}}\sim\exp(-2\Phi')$], the reflectance has considerably deeper minima [cf. Fig.~\ref{[10]-}~(a)
and Fig.~\ref{[10]+}~(c)], in accordance with the above-mentioned general property of superstrate resonances. The
amplitudes of nonzeroth-order outgoing transmitted waves exceed significantly the zeroth order amplitude [compare the
solid curve for $\tau_{\mathcal{O}}$ and the dotted curve for $\tau_{[1,0]}$ in Fig.~\ref{[10]-}~(a)] since their
amplitudes are of order $\tau_{[1,0]}\sim\xi_{\mathcal{O}}'^{-1}\exp(-2\Phi')$. It follows from the Eq. \eqref{b1}.
This estimation is quite universal and coincide with that made using the expression Eq.~\eqref{ss1} for the simplest
resonance. The enhanced nonzeroth-order transmission may be used for developing the light-splitters, since the waves
propagate at an angle $\mathrm{arccos}(\beta_{r|+}\sqrt{\varepsilon_+})$ relative to $z$ axis. The unique feature of
such a splitter is that all the four nonzeroth-order waves are linear polarized. Their polarization coincides with that
of SPPs excited.\cite{KNN_PhysRev}

Note that not only $[1,0]$ diffraction orders are responsible for the nonzeroth-order ELT. As seen from
Fig.~\ref{[10]-} (b), the diffraction orders $[1,1]$ correspond to the propagating waves in the quartz substrate as
well. The resonance scattering contributes to these amplitudes amplitudes so that they are proportional to
$\tilde{\xi}_{[1,0]}^2$. Their intensities have been taken into account when calculating the total transmitted energy
flux shown by the dashed line in Fig.~\ref{[10]-}~(a).

\begin{figure}[!bt]
  \includegraphics[width=7cm]{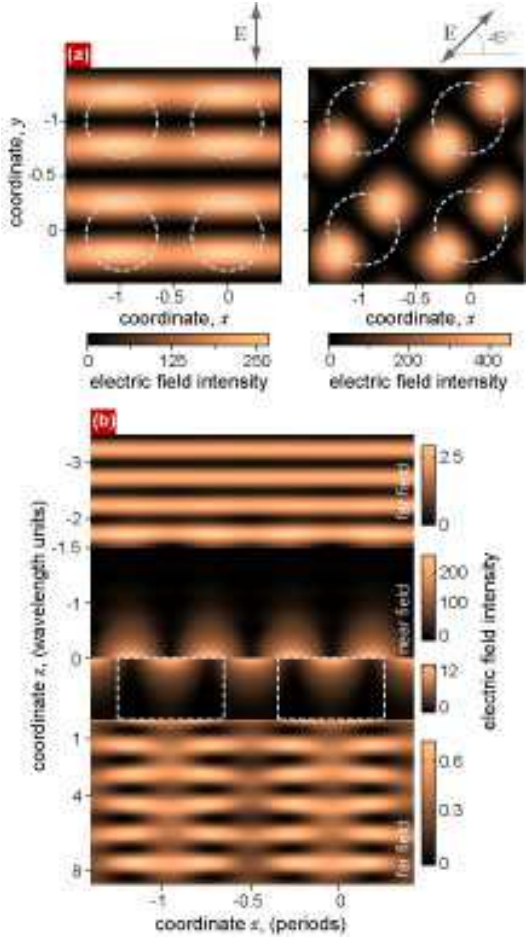}\\
  \caption{\label{[10]-field}
The intensity distribution is given for $[1,0]_-$ SPP resonance at
$\lambda=0.9234~\mu m$, $\theta=0$. The parameters of the array are
the same as in Fig.~\ref{All_NS}.}
\end{figure}

It has to be pointed out that the transmittance behavior with respect to the polarization of the incident wave change,
in the vicinity of $[1,0]_-$ resonance, is similar to that in the vicinity of $[1,0]_+$ resonance. This is described in
Subsection~\ref{subsec:[10]+N}. (cf. the dashed square regions of $\lambda-\theta$ and the upper solid square regions
in Fig.~\ref{t2NS}.) The near-field distribution in $X-Y$ plane is likewise equivalent to that corresponding to
$[1,0]_+$ resonance, but now the field is localized at the air-metal interface. This is clearly seen in
Fig.~\ref{[10]-field}~(b). Note that the far-field structure has the form of the interference pattern both in the air
and in the quartz half-spaces. The pattern in the air superstrate is due to the interference between incident and
reflected zeroth-order waves: we see from Fig.~\ref{[10]-}~(a) that the reflectance is of order of 20\% in the
resonance. The pattern in the quartz substrate is caused by the interference between the transmitted zeroth-order and
nonzeroth-order waves.

Let us illustrate the dependence of the zeroth and non-zeroth order transmittance upon the modulation amplitude. The
transmittance (reflectance) has the well-defined maximum at $\xi_i\simeq2\xi_f$, which corresponds to
$|\tilde{\xi}_{[1,0]}|\simeq0.6\sqrt{\xi_{\mathcal{O}}'}\sim \sqrt{\xi_{\mathcal{O}}'}$, see Fig.~\ref{[10]-}~(c). So,
as expected, this value of the modulation amplitude guarantee exactly the resonance harmonic optimal value, see
discussion below Eq.~\eqref{ss7}, and, therefore, leads to the optimal transmittance value.

\subsection{\label{subsec:NsymmDBSPP}Double bounday SPP}

DB SPPs for nonsymmetrically-surrounded film exist under a specific relation between angle of incidence, wavelength,
structure spacing, and dielectric permittivities of the surrounding media\cite{film_PhysRev}. For instance, in the case
of strictly normal incidence the excitation of four resonance waves in the superstrate ($\tau=+$) having
multiindexes $[r_+,0]$, and four resonance waves in the substrate ($\tau=-$) having multiindexes $[r_-,0]$
may take place if $K_+/K_-=r_+/r_-$. We do not consider this specific set of parameters in the present paper, but  study DB SPPs in a symmetrically-surrounded film below.

\section{\label{sec:symmetric}DB SPP resonances in the symmetrically-surrounded film}

To the best of our knowledge, the earliest observations  of the ELT through hole arrays due to the DB SPP excitation
were performed in Refs.~\onlinecite{Ebbesen2001_Opt,Ebbesen_theory_PRL01}. Moreover, in
Ref.~\onlinecite{Ebbesen2001_Opt} the array was formed in the metal film surrounded by quartz from one face and by a
liquid with a close-to-quartz dielectric constant from the other face, while Ref.~\onlinecite{Ebbesen_theory_PRL01}
deals with a free-standing film. In the above experiments the matching of wavevectors at the opposite interfaces was
made to excite a DB SPP. This considerably increased the ELT efficiency in zeroth diffraction order as compared to the
SB SPP excitation for a nonsymmetrically surrounded film.

As known, there exist two SPP modes in the nonmodulated symmetrically-surrounded metal film, see
Sec.~\ref{sec:general}. One of them, a LR mode, corresponds to the nonsymmetric surface charge distribution, and higher
frequency and Q-factor as compared to the other, SR mode.  A remarkable feature is that the similar situation is
inherent to the mechanical system of two coupled damping oscillators which is nothing else but a classical analog of
coupled SPPs in the film, see Appendix~\ref{oscillators}. If the modulation is not too high, LR and SR modes undergo
shift and widening, but remain well-separated from one another. This is not exactly what is observed in the experiments
using subwavelength hole arrays. The modulation amplitude of the arrays is too high to distinguish well the difference
between LR and SR modes.

\begin{figure}[!tb]
  \includegraphics[width=8cm]{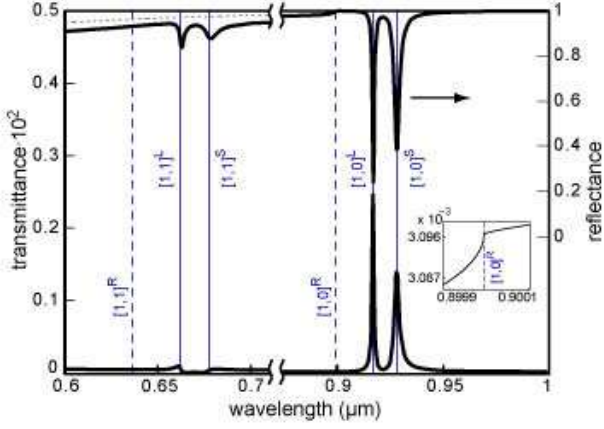}\\
  \caption{\label{All_S}(Color online)  Dependence of
transmittance and reflectance upon the wavelength for strictly normal incidence onto the symmetrically surrounded (by
air) silver film. The periodical array possesses $C_{4v}$ symmetry, i.e. $\rho_1=\rho_2=\rho=0.9\mu m$, $a/\rho=1/3$.
$\mathbf{E}$ is parallel to $\mathrm{\mathbf{g}}_1$ (to $x$ axis, see. Fig.~\ref{fig:geometry}). The impedance of
inclusions, $\xi_i$, is assumed to be $\xi_i=2\xi_f$. The film thickness is equal to $2.75$ skin depths. The
transmittance and reflectance corresponding to the unmodulated film are shown by dotted lines. Vertical solid (dashed)
lines indicate the wavelengths for SPPs at the unmodulated boundaries (Rayleigh points), one of them is shown in the
inset.}
\end{figure}

\begin{figure*}[!htb]
  \includegraphics[width=17cm]{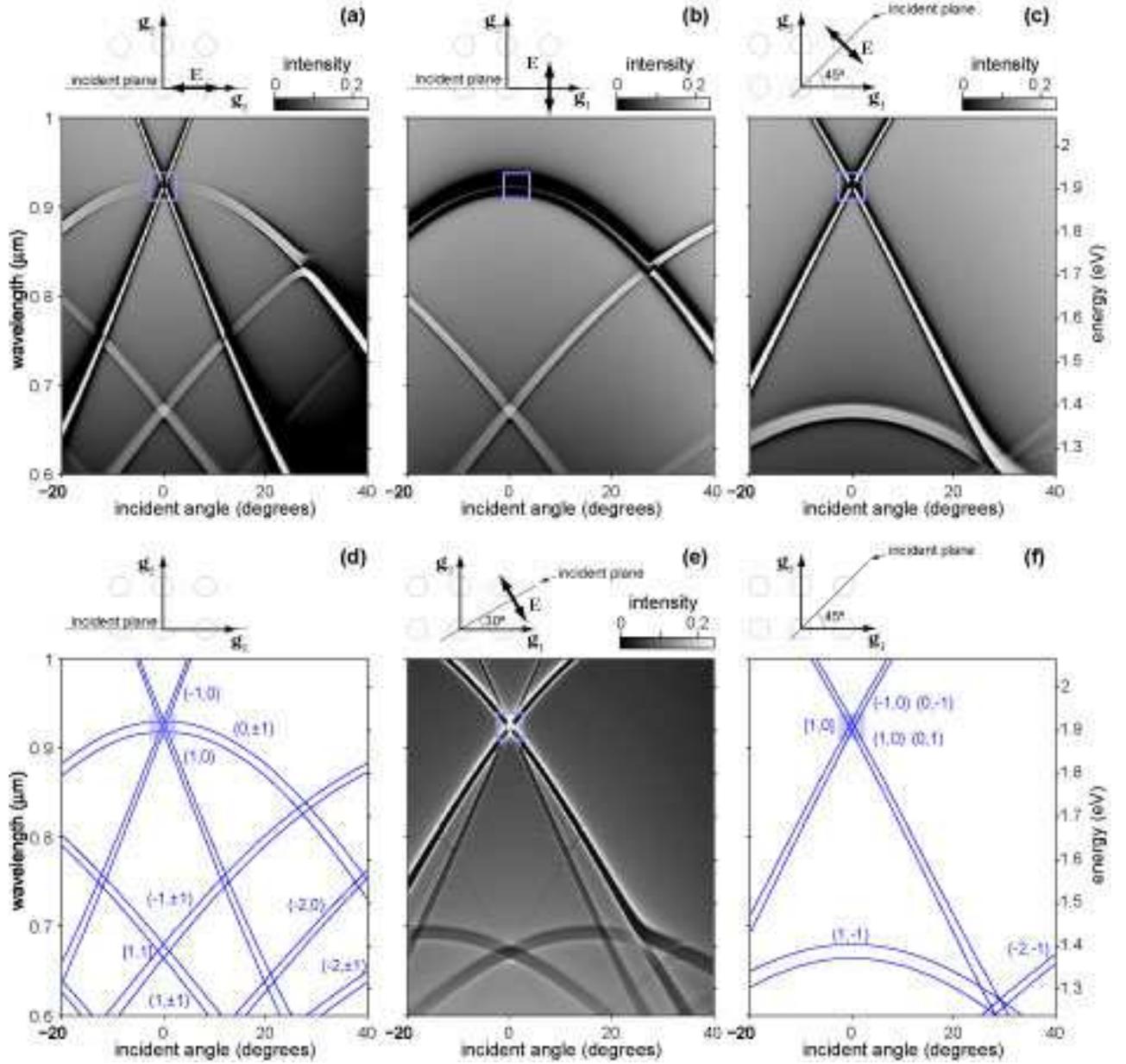}\\
  \caption{\label{t2S}(Color online) Dependence of
transmittance upon the wavelength (energy) and the incident angle for different
polarization of the incident wave and different orientations of the incident plane. The parameters of the array and the film are the same as in Fig.~\ref{All_S}. (d, f) the resonance curves [see Eq.~\eqref{45.2.0}]. Squares indicate the vicinity of the $[1,0]$ resonance which will be considered in detail below.}
\end{figure*}

The transmittance spectra of a symmetrically-surrounded film is shown in Fig.~\ref{All_S}. One can see the four peaks:
the first pair is $0.663~\mu m$, $0.683~\mu m$ in the vicinity of $[1,1]^{S,L}$ DB SPP resonance, and the second pair
is $0.918~\mu m$, $0.929~\mu m$ in the vicinity of $[1,0]^{S,L}$ DB SPP resonance. Indexes $S$, $L$ comply with LR and
SR modes. In Figs.~\ref{t2S} (a-c) and (e) the transmittance spectra are presented as functions of the wavelength
and/or energy and the incident angle for different polarizations of the incident wave. In Figs.~\ref{t2S} (d,f) the
resonance curves are shown by solid lines.

Consider the deviations  from the normal incidence. In Figs.~\ref{t2S}  (a-c) the special symmetry is shown, that is,
the incident plane is oriented relative to $\mathbf{g}_1$ at $0$ and $\pi/4$. This complies with coincidence of the
resonance curves, see discussion below formula \eqref{45.2.0}. It is seen in the vicinity of $[1,0]$ resonance shown by
square regions in Fig.~\ref{t2S}. In Figs.~\ref{t2S} (a-c) from six to two resonance features belong to these regions.
These features correspond from six to two resonance curves in Figs.~\ref{t2S} (d,f). Oppositely, in Fig.~\ref{t2S}  (e)
the transmittance contour plot is shown for non-specific geometry, and therefore, eight features (or, eight curves)
belong to each vicinity according to fourfold DB SPP.

Note the two features in Fig.~\ref{All_S}. First, the widths of
resonance peaks (dips) are less for LR modes than for SR modes. This is due to the higher
Q-factor of LR mode as compared with SR mode, that can be easily explained in terms of
field structure of SPP modes in the metal. $z$-dependence of the electric field
tangential component for LR and SR eigenmodes is defined by Eq.~\eqref{mr3}.
Since the amplitude of the tangential component of the electric field is higher than the
$z$-component within the film, vector $\overline{\mathbf{E}}_t$ makes the main contribution to the loss power,
$P$, so that
\begin{equation}\label{pls}
\mathcal{P}\sim\int\varepsilon''|\overline{\mathbf{E}}|^2dv\sim
\int_{0}^{d}|\overline{\mathbf{E}}_t|^2dz,
\end{equation}
In view of Eq.~\eqref{mr3}, one can see that the Ohmic losses of SR mode is higher than that of LR mode,
$\dfrac{P_s}{P_l}\sim\dfrac{\sinh\Phi'-\Phi'}{\sinh\Phi'+\Phi'}>1$, which leads to the higher Q-factor of LR mode as
compared with SR mode.

Second, the resonance peaks for LR and SR modes is strongly dependent upon the exponential index appearing in the field
approximation inside the film. In other words, assuming that $\tilde{k}$ in Eq.~\eqref{3} is complex and taking into
account both real and imaginary parts of the impedance of the conductor in Eq.~\eqref{3} the transmittance maxima
(reflectance minima) for LR mode are higher (deeper) than those for SR mode. And vice versa if dissipation losses are
neglected. As was indicated in the experimental paper\cite{ener.transfer_science04}, where plasmons were studied in the
symmetrically surrounded corrugated silver film, ``this process has a complex distance (film thickness) dependence'',
that is ``for thin silver films ($<30$ $nm$), the symmetric coupled SPP is strong but extremely sharp, whereas the
antisymmetric SPP is broad but weak. As the silver thickness increases, the symmetric mode weakens and broadens,
whereas the antisymmetric mode sharpens and intensifies''. Thus, to our opinion, the discussed question of the
resonance peaks for LR and SR modes needs further careful studying.

\subsection{\label{subsec:[10]S}$[1,0]$ resonance}

\begin{figure*}[!htb]
  \includegraphics[width=17cm]{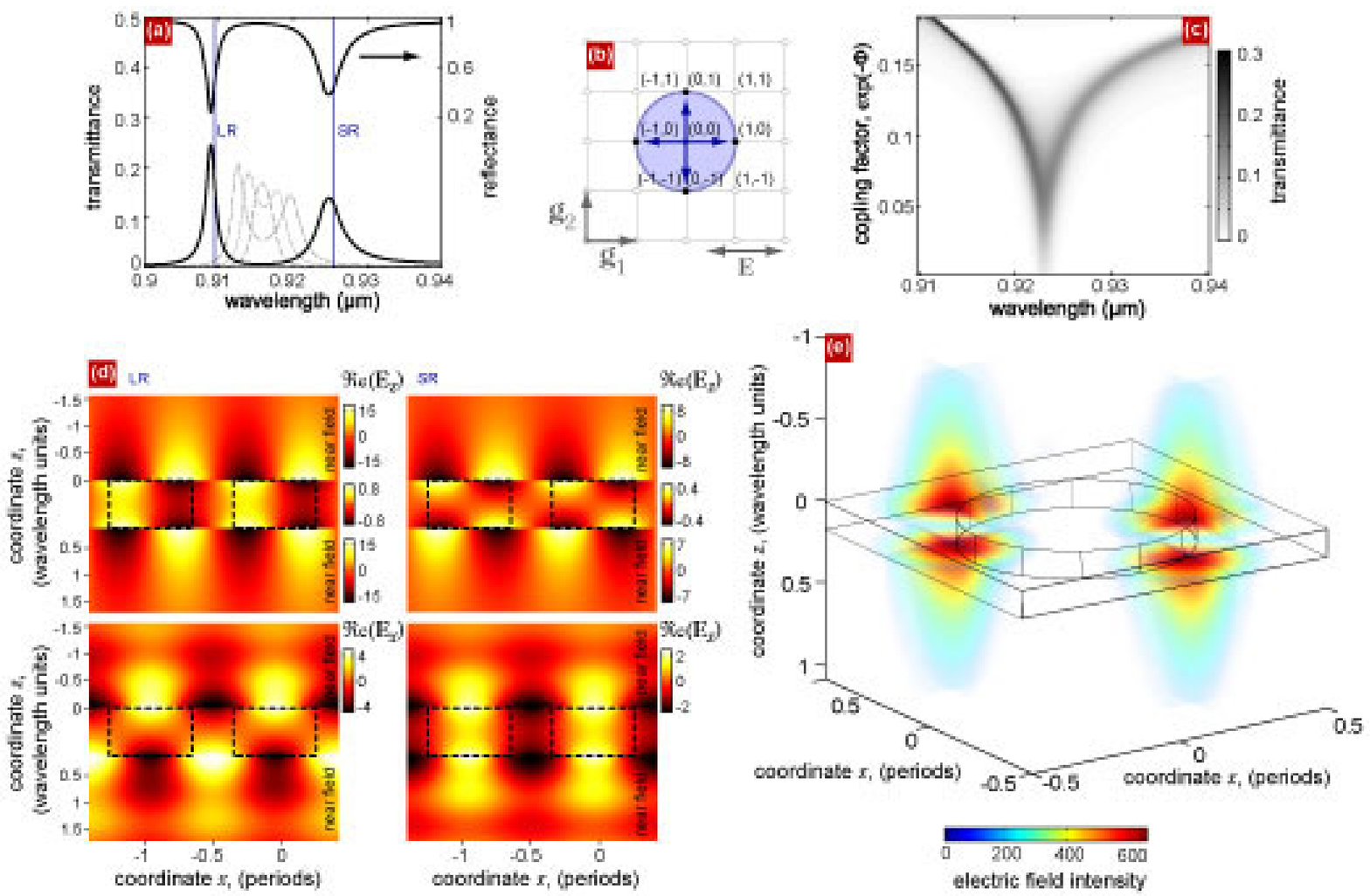}\\
  \caption{\label{[10]} (Color online) (a) The dependence of the zeroth-order
transmittance, $\tau_{\mathcal{O}}$, and reflectance, $\rho_{\mathcal{O}}$, upon the wavelength corresponding to
$[1,0]$ resonance on the silver array of inclusions (with the symmetry $C_{4v}$), $\mathbf{E} \| \mathbf{g}_1$, $\theta=0$. (b) the reciprocal grating [compare with Figs.~\ref{[10]+} (b),~\ref{[11]+} (b),~\ref{[10]-}
(b)]. (c) the dependence of $\tau_{\mathcal{O}}$ upon the wavelength and film thickness for $\theta=0$.
(d, e) the distribution of tangential and normal components of the electric field at $\lambda=0.918 \mu m$
(for LR mode) and $\lambda=0.9285 \mu m$ (for SR mode), $\theta=0$. The inclusions in (d) are shown by the dashed
lines. The parameters of the array are the same as in Fig.~\ref{All_S}.}
\end{figure*}

Now consider the $[1,0]$ DB fourfold SPP resonance for  strictly normal incidence. The enlarged fragment of
Fig.~\ref{All_S} in the vicinity of the peaks is shown in Fig.~\ref{[10]} (a). The LR and SR resonances are blueshifted
as compared with the wavelengths for nonmodulated film SPPs marked by the vertical lines. Analogously to the SB SPP
resonance, the amplitudes of the resonance waves are proportional to $\tilde{\xi}_{[10]}$, and the resonance
contribution to the amplitude of the only propagating zeroth-order wave is proportional to $|\tilde{\xi}_{[10]}|^2$.

For a SDB resonance, the transmittance and reflectance have two extremums. This results from the excitation of LR and
SR SPPs, corresponding to the zeros of the denominators $\Delta_{[1,0]}$ in Eqs.~\eqref{ss6}, \eqref{ss7} (and
Appendix~\ref{C2v}). The peak value of the transmittance for optimal modulation amplitude may be estimated as
$\tau_\mathcal{O}\sim\xi_{\mathcal{O}}'^{-1}\exp(-\Phi')$, which exceeds the transmittance through the unmodulated film
by a factor of $\xi_{\mathcal{O}}'^{-1}\gg1$.

The distance between the two extremums of the transmittance (reflectance) may be estimated by extracting the difference
between the wavelengths corresponding to  SR (LR) SPPs from the dispersion relations Eq.~\eqref{45.2.1}. Using
$\beta_{\mathcal{M}}^{L,S}=\varepsilon^{-1/2}\sqrt{1-(cq/\sqrt{\varepsilon}\omega^{L,S})^2}$, (where we have replased
$\boldrm{k}_{\mathcal{M}t}$ by the SPP wavevector, $\mathbf{q}$), we find for the frequencies $\omega^{L,S} =
cq/\sqrt{\varepsilon B^{L,S}}$, where for the  LR mode $B^{L}=1-\varepsilon\xi_{\mathcal{O}}^2\tanh^2(\Phi/2)$ and for
the SR mode $B^{S}=1-\varepsilon\xi_{\mathcal{O}}^2\coth^2(\Phi/2)$. In terms of wavelength we have for the relative
difference between the wavelengths $(\lambda^{S}-\lambda^{L})/\lambda\sim|\xi_{\mathcal{O}}''|^2\exp(-\Phi')$, where
$\lambda$ is of order of $\lambda^{S}$, $\lambda^{L}$. The modulation contributes additionally to this difference, and
this contribution may be found from the dispersion relation $\Delta_\mathcal{R}=0$, Eq.\eqref{ss4}, see
Ref~\onlinecite{film_PhysRev}. As the film thickness increases, the distance between $\lambda^{S}$ and $\lambda^{L}$
decreases, see Figs.~\ref{[10]} (a), (c). When the resonance width,
$\Delta\lambda/\lambda\sim|\xi_{\mathcal{O}}''|\xi_{\mathcal{O}}'$, takes the value of order of the splitting between
the LR and SR modes, $\Delta\lambda\sim\lambda^{S}-\lambda^{L}$ [which occurs for film thickness
$\Phi'\gtrsim\ln(|\xi_{\mathcal{O}}''|\xi_{\mathcal{O}}'^{-1})$], the two maxima of the two-humped resonance curve
become indistinguishable.

The near-field distribution for LR and SR SPPs is shown in Fig.~~\ref{[10]} (d), (e). One can see that LR (SR) mode is
antisymmetric (symmetric) with respect to the tangential component of the electric field and the surface charge
distribution (the surface charge sign coincides with the sign of the production  $n_zE_z$, where $n_z$ is $z$-component
of the surface normal vector pointing out of the metal).

\section{\label{conclusions}Conclusions}

We have treated analytically the resonance optical properties of $2D$ periodically-modulated optically-thick metal
films.  Explicit analytical expressions for transformation coefficients related to any diffraction order have been
obtained. In studying the complicated multiple resonances, we have ascertained that the most of the physical properties
of ELT may be well understood in terms of the simplest example of a solitary resonance (when SPP is excited in a single
diffraction order). We have examined and explained not only the amplitude and polarization dependencies of the
transmittance and reflectance upon the parameters (angle of incidence, wavelength, tilting angle, film thickness,
etc.), but also the field structure of the diffracted light. We have shown that according to the conception of
interface SPPs excitation, the polarization dependencies of the light diffracted by the periodical array may be
adequately described. We have compared our theoretical calculations with recent experiments and found an excellent
coincidence. We have shown that for hole arrays (which are in most cases used for the experimental study of ELT) it is
difficult to make a distinction between the long-range and short-rang SPP modes excitation caused resonance features
since the radiative broadening exceeds the splitting between them for rather thick films. In this sense structures with
weak modulation (say, the periodical arrays of metal nanoparticles immersed into a metal film, corrugated shallow
diffraction gratings) are more preferable. Moreover, the efficiency of SPP excitation may be much higher for the
structures with weak modulation as compared to the structures with holes being included. This is due to the existence
of the optimal modulation amplitude. It was demonstrated that in some cases the energy flux corresponding to nonzeroth
diffraction orders, exceeds that for zeroth-order. This can have a strong influence upon the results of experimental
measurements.

A highly instructive analogy was drawn between motion of the monochromatic force driven system of two weakly coupled
classical oscillators and ELT effect.

\section{\label{conclusions}Acknowledgements}

The authors acknowledge financial support from the INTAS YS grants Nr 05-109-5206 and Nr 05-109-5182.

\appendix

\section{\label{oscillators}Two coupled damping oscillators}

The enhanced transparency and other resonant properties of the periodically modulated metal film can be described by
analogy with a simple mechanical system consisting of two weakly coupled harmonic oscillators under the action of the
harmonic force applied to one of them see Fig.~\ref{maS} (a).
\begin{figure*}[!tb]
  \includegraphics[width=17cm]{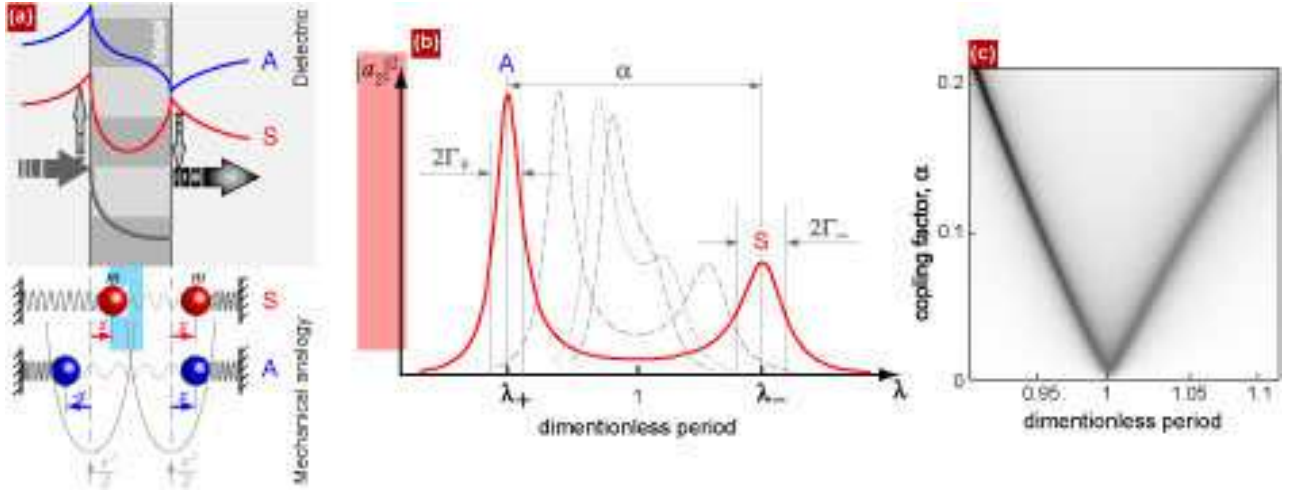}\\
\caption{\label{maS}(Color online) (a) The geometry of the problem and  the mechanical analogy. (b) The dependence of
``free'' oscillator amplitude upon the period, $1/\Om$. The coupling parameter and the damping are: $\a=0.05$, $0.03$,
$0.015$, $0.01$ for the solid,  dash-dotted, dotted, dashed curve respectively; $\G_1=0.004$, $\G_2=0.001$ for all
curves.  (c) Dependence of the ``free'' oscillator amplitude on the period, $1/\Om$, and coupling parameter
$\a$.}\label{maS}
\end{figure*}
An oscillation amplitude of the second oscillator is strongly dependent on the frequency of the force applied to the
first oscillator  and can reach amplitudes of order of the amplitude of the first uncoupled oscillator. We will
concentrate on the system of two identical oscillators which has the analogy with the ELT through
symmetrically-surrounded film. Lagrangian of this system is
\begin{eqnarray}\label{o1}
    \mathcal{L}=\frac{1}{2}\left(\dot{x_1}^2+\dot{x_2}^2\right)-
    \frac{1}{2}\left(x_1^2+x_2^2\right)+ \a x_1x_2\nonumber\\
     + F_{1}(t)x_{1} + F_{2}(t)x_{2} \,,
\end{eqnarray}
where dots denote time derivatives, the term $\a x_1x_2$ is responsible for weak coupling, $\a \ll 1$, and the two last
terms are for the forces acting on the oscillators. Adding to the Lagrangian $-k (x_1^2+x_2^2)/2$ we arrive at the two
masses linked by a string with elasticity coefficient $k$.  Units are such that the unperturbed frequency equals unity.
The losses are taken into account by the dissipative function\cite{Landau},
\begin{equation}\label{o2}
    \mathcal{F} =
   \mu_1(\dot{x}_1^2+\dot{x}_2^2)+  2\mu_2\dot{x}_1\dot{x}_2 \,,
\end{equation}
where $\mu_1 \ge |\mu_2| \ge 0$ (non-negativeness of $\mathcal{F}$). Below the
dissipation is supposed to be small, $|\mu_{1,2}| \ll 1$. Then equations of motion are
\begin{equation}\label{o3}
    \frac{d}{d t}\frac{\p \mathcal{L}}{\p \dot{x}_i} =
    \frac{\p \mathcal{L}}{\p x_i} - \frac{\p \mathcal{F}}{\p \dot{x}_i} \,
    , \quad i = 1,2 .
\end{equation}

In normal coordinates, $X_{\mp} = (x_{1} \pm x_{2})/\sqrt{2}\,$, which correspond to the
inphase ($X_-$) and antiphase ($X_+$) oscillations (symmetric and antisymmetric with
respect to transposition $1\rightleftarrows2$ modes), we split the system \eqref{o3},
\begin{equation}\label{o3.1}
    \ddot{X}_{\mp}+(1 \mp  \a)X_{\mp} + 2 (\mu_1\pm\mu_2)\dot{X}_{\mp} = F_{\mp} \, ,
\end{equation}
where $F_{\mp} = (F_{1} \pm F_{2})/\sqrt{2}\,$.

Solution of the corresponding homogeneous system is
\begin{equation}\label{o4}
    X_{\pm}=A_{\pm}\exp(-i\Omega_{\pm}t) \, ,
\end{equation}
where the complex eigenfrequencies are $\Omega_{\mp}$ and $-\Omega_{\mp}^{\ast}$ ($\Re e
\Omega_{\mp}> 0$, $\Im m\Omega_{\mp}\leq0$),
\begin{eqnarray}\label{o5}
\Om_{\mp}=-i(\mu_1\pm\mu_2) + \sqrt{1\mp  \alpha - (\mu_1\pm\mu_2)^2} \nonumber\\
\simeq 1
\mp \frac{\alpha}{2} - \frac{(\mu_1\pm\mu_2)^2}{2} -i(\mu_1\pm\mu_2 )  \, .
\end{eqnarray}
Eigenfrequency of the inphase (symmetric) mode is lower than that for antiphase (antisymmetric) one, $\Re e \Omega_{-}
< \Re e \Omega_{+}$ if the damping is small, $(\mu_1\pm\mu_2)^2 \ll \alpha$. This is in compliance with the  modes of
the symmetrically-sandwiched film. Namely, eigenfrequency of the LR (antisymmetric) SPP is lower than that for SR
(symmetric) one, see discussions in the Subsection~\ref{subsec:[10]S}.  In turn, the corresponding decrement, $\G_{\mp}
\simeq \mu_1\pm\mu_2$, is higher (lower) for $\mu_2 > 0$ ($\mu_2 < 0$). Note that the case with $\mu_2 > 0$ holds for
different active losses for symmetric and antisymmetric polariton modes. This difference is due to the fact that the
high-frequency mode has a lower mean value of the electric field in the film, and, hence, lower losses and a higher
$Q$-factor.

The specific solution of the dynamic equations~\eqref{o3.1} for harmonic driving forces
being applied, $F_{1,2} = f_{1,2} \exp(-i\Omega t)$, is
\begin{equation}\label{o6}
    X_{\pm}(t) = A_{\pm} (\Omega) \exp(-i\Omega t) , \quad A_{\pm} =
    f_{\pm}/\mD_{\pm} \, ,
\end{equation}
where $f_{\pm} = (f_{1} \mp f_{2})/\sqrt{2}$,
\begin{equation}\label{o7}
    \mD_{\pm} =     \mD_{\pm}(\Omega ) =-(\Om-\Om_{\pm})(\Om+\Om^{\ast}_{\pm}).
\end{equation}
Thus, displacements of the forced oscillations are $x_{i}(t) = a_{i}(\Omega)
\exp(-i\Omega t)$, $i = 1,2$,
\begin{equation}\label{o8}
a_{1,2}(\Omega) = \frac{A_{-} \pm A_{+}}{\sqrt{2}} = \frac{f_1 + f_2}{2 \mD_-} \pm
\frac{f_1-f_2}{ 2 \mD_+}  .
\end{equation}

Let excitation of the system be caused by a driving force applied to the first oscillator,  $f_1 = 4$, $f_2 = 0$. This
is analogous to the diffraction problem, where only one wave is incident onto a face of the metallic film.
 If
the dissipation is small, $(\mu_1\pm\mu_2)^{2} \ll \alpha$, then there exist two pronounced maximal magnitudes of
$a_{1,2}$ in the vicinity of the frequencies $\Omega = \Omega_{\mp}' $, where $a_{1,2}(\Omega_{-}') \simeq
2/\mD_{-}(\Omega_{-}') \simeq i/\Gamma_{-}  $, $a_{1,2}(\Omega_{+}') \simeq \pm 2/\mD_{+}(\Omega_{+}') \simeq \pm
i/\Gamma_{+}  $. The splitting of the resonance maxima is approximately $\a$, and the up- (low-) frequency resonance
widths are $\delta \Omega_{\pm} = 2\Gamma_{\pm}$. As $\a$ decreases, the distance between resonance peaks diminishes,
and for $\a \sim \G_{+} + \G_{-}$ they overlap, see Fig.~\ref{maS} (b), (c). The mechanical coupling factor $\alpha$ is
analogous to that arising in the EM problems: for the splitting of the SPP modes in the film this factor is $\exp
(-\Phi')$, compare Fig.~\ref{maS} (c) and Fig.~\ref{[10]} (c).

 For $\mu_{1} = \mu_{2} =0$ the amplitude
of the enforced oscillator vanishes at the unperturbed eigenfrequency, $\Omega = 1$. This well-known effect of the
resonance damping (which is extensively used in shipbuilding) also shows an analogy with the ELT problem, where Fano
minima arise in the amplitudes of the fields.

In a similar way one can draw an analogy for the nonsymmetrically-sandwiched film. In  that case the mechanical system
consist of two oscillators of different masses.

Thus, some physical properties of the ELT phenomena are quite general. At least, the similarities with the simplest
classical mechanical systems may be established.

\section{\label{C2v} Transformation coefficients for fourfold resonance in
cases of $C_{2v}$ and $C_{4v}$ symmetries}

In the case of $C_{2v}$ symmetry which implies the following properties of modulation
impedance:
$$
\tilde{\xi}_{n_1,n_2}=\tilde{\xi}_{-n_1,n_2}=\tilde{\xi}_{n_1,-n_2}=\tilde{\xi}_{-n_1,-n_2},
$$
the resonance matrix possesses the symmetry properties.  The resonance TCs in the vicinity of normal incidence for
$[r,0]$, or $[r,0]_\tau$ resonance are related as $T^{\tau|-\sigma}_{r,0}=-T^{\tau|-\sigma}_{-r,0}$,
$T^{\tau|-\sigma}_{0,r}=-T^{\tau|-\sigma}_{0,-r}$. This leads to the essential simplification of Eqs.~\eqref{r},
\eqref{r1}: the resonance matrix $8\times8$ can be reduced to a matrix $2\times2$. Thus, the TCs are written as
\begin{eqnarray}\label{b1}
&&T_{\mathcal{R}}^{\tau|-\sigma} =  -\frac{2\tau\sigma
\sqrt{\varepsilon_-}S_{\mathcal{R}\mathcal{O}}^{\overline{\sigma}}
\tilde{\xi}_{\mathcal{R}}\beta_{-|\mathcal{O}}^{\frac{1+\sigma}{2}}}{\Delta_{\mathcal{R}}}
\nonumber\\
&&\times\left[\tilde{\beta}_{\overline{\tau}|\mathcal{R}}-
\Upsilon_{\mathcal{R}}(\mathrm{cosh}\Phi)^{\tau-1}\right](\mathrm{cosh}\Phi)^{-\frac{1+\tau}{2}}\; ,
\end{eqnarray}
where $\mathcal{R}=(r,0)$, or $\mathcal{R}=(0,r)$,
\begin{eqnarray}\label{b2}
\Delta_{\mathcal{R}}= \tilde{\beta}_{+|\mathcal{R}}\tilde{\beta}_{-|\mathcal{R}} -
\Upsilon_{\mathcal{R}}^{2}\mathrm{cosh}^{-2}\Phi ,\nonumber\\
\tilde{\beta}_{\tau|\mathcal{R}}\equiv \beta_{\tau|\mathcal{R}}\tanh\Phi + \xi_{\mathcal{O}}+\tilde{\xi}_{2\mathcal{R}}
+ G_{\mathcal{R}}^{\tau},\nonumber\\
\Upsilon_{\mathcal{R}}=\xi_{\mathcal{O}}+\tilde{\xi}_{2\mathcal{R}}+
G_{\mathcal{R}}^{+}+G_{\mathcal{R}}^{-},\nonumber\\
G_{\mathcal{R}}^{\tau}=-\sum\limits_{\mathcal{N}}\dfrac{\tilde{\xi}_{\mathcal{R}-\mathcal{N}}\left(
\tilde{\xi}_{\mathcal{N}-\mathcal{R}}+
\tilde{\xi}_{\mathcal{N}+\mathcal{R}}\right)}{\beta_{\tau|\mathcal{N}}}\nonumber\\
\times(C_{\mathcal{R}\mathcal{N}}^2+\varepsilon_{\tau} \beta_{\tau|\mathcal{N}}^2S_{\mathcal{R}\mathcal{N}}^2),
\end{eqnarray}
%\begin{widetext}
\begin{eqnarray}\label{b3}
T^{\tau|\sigma\sigma'}_{\mathcal{N}} \!\!\!=
\delta_{\mathcal{N},\mathcal{O}}\delta_{\sigma,\sigma'}T_{F}^{\tau|\sigma}+
A^\tau_{\mathcal{N}|\sigma} (\mathrm{cosh}\Phi)^{-\frac{1+\tau}{2}}\nonumber\\
\times\sum_{\mathcal{R}} \frac{\tilde{\xi}_{\mathcal{R}}S_{\mathcal{N}\mathcal{R}}^{\overline{\sigma}}
S_{\mathcal{R}\mathcal{O}}^{\overline{\sigma'}}
\left(\tilde{\xi}_{\mathcal{N}-\mathcal{R}}+\tilde{\xi}_{\mathcal{N}+\mathcal{R}}\right)}
{\Delta_{\mathcal{R}}}\nonumber\\
 \times\left[\tilde{\beta}_{\overline{\tau}|\mathcal{R}}+
(\tilde{\beta}_{\tau|\mathcal{R}}-\Upsilon_{\mathcal{R}})(\mathrm{cosh}\Phi)^{\tau-1}\right] ,
\end{eqnarray}
%\end{widetext}
where
$A^\tau_{\mathcal{N}|\sigma}=2\sigma'\tau^{\frac{1-\sigma}{2}}\varepsilon_\tau^{\frac{1+\sigma}{4}}
\varepsilon_-^{\frac{1+\sigma'}{4}}\beta_{-|\mathcal{O}}^{\frac{1+\sigma'}{2}}
\beta_{\tau|\mathcal{N}}^{\frac{\sigma-1}{2}}$. If the modulation has  $C_{4v}$
symmetry, the zeroth-order TCs allow additional simplifications
%\begin{widetext}
\begin{eqnarray}\label{b4}
T^{\tau|\sigma\sigma}_{\mathcal{O}} =&& T_{F}^{\tau|\sigma}+ \frac{2\sigma
A^\tau_{\mathcal{O}|\sigma}\tilde{\xi}_{\mathcal{R}}^2}
{\Delta_{\mathcal{R}}}\nonumber\\
&&\times \left[\tilde{\beta}_{\overline{\tau}|\mathcal{R}}+
(\tilde{\beta}_{\tau|\mathcal{R}}-\Upsilon_{[r,0]})(\mathrm{cosh}\Phi)^{\tau-1}
\right]\nonumber\\
&&\times (\mathrm{cosh}\Phi)^{-\frac{1+\tau}{2}}, \q
T^{\tau|\sigma\overline{\sigma}}_{\mathcal{O}}=0.
\end{eqnarray}
%\end{widetext}
 Notice that the zeroth-order TCs are diagonal in polarization, i.e. the polarization of zeroth-order
 reflected/transmitted wave is the same as that of the incident wave.


\begin{thebibliography}{99}
%1
\bibitem{Book_Raether} H.~Raether, {\it Surface plasmons}, (Springer-Verlag, New York, 1988).
%2
\bibitem{Book_Agranovich} V.~M.~Agranovich and D.~L.~Mills, {\it
Surface Polaritons}, (Nauka, Moscow,  1985).
%3
\bibitem{Zayats_Maradudin_2005} A. V. Zayats, I. I. Smolyaninov,
A. A. Maradudin, Phys. Rep. \textbf{408},  131 (2005)
%4
\bibitem{Ebbesen_Nature_03} W.~L.~Barnes, A.~Dereux, and
T.~W.~Ebbesen, Nature \textbf{424}, 824 (2003).
%5
\bibitem{Ebbesen98_Nature} T.~W.~Ebbesen,  H.~J.~Lezec,
H.~F.~Ghaemi, T.~Tio, and P.~A.~Wolff, Nature \textbf{391}, 667 (1998).
%6
\bibitem{Ebbesen_PRB_98} H.~F.~Ghaemi, T.~Thio, D.~E.~Grupp,
T.~W.~Ebbesen, and H.~J.~Lezec, Phys. Rev. B. \textbf{58}, 6779 (1998).
%7
\bibitem{Bethe} H. A. Bethe,  Phys. Rev. \textbf{66}, 163 (1944).
%8
\bibitem{Ebbesen2001_Opt} A.~Krishnan, T.~Thio, T.~J.~Kim,
H.~J.~Lezec, T.~W.~Ebbesen, P.~A.~Wolf, J.~Pendry, L.~Martin-Moreno and F.~J.~Garc\'{i}a-Vidal, Opt. Com. \textbf{200},
1 (2001).
%9
\bibitem{Ebbesen_theory_PRL01} L.~Mart\'{i}n-Moreno,
F.~J.~Gars\'{i}a-Vidal, H.~J.~Lezec, K.~M.~Pellerin, T.~Thio, J.~B.~Pendry, and T.~W.~Ebbesen, Phys. Rev. Lett.
\textbf{86}, 1114 (2001).
%10
\bibitem{StrongCoupling_PRB05} J. Dintinger, S. Klein, F. Bustos, W. L. Barnes, and T. W. Ebbesen,
 Phys. Rev. B. \textbf{71}, 035424 (2005).
%11
\bibitem{Fluorescence_OptLett_03} Y.~Liu, and S.~Blair, Opt.
Lett. \textbf{28}, 507 (2003).
%12
\bibitem{Barnes_PRB04} W.~A.~Murray, S.~Astilean, and W.~L.~Barnes, Phys. Rev. B \textbf{69}, 165407 (2004).
%13
\bibitem{hole_depth_APL02} A.~Degiron, H.~J.~Lezec, W.~L.~Barnes, and T.~W.~Ebbesen, Appl. Phys. Lett. \textbf{81}, 4327
(2002).
%14
\bibitem{shadows_APL_02} S.~C.~Hohng, Y.~C.~Yoon, D.~S.~Kim,
V.~Malyarchuk, R.~M\"{u}ller, Ch.~Lienau, J.~W.~Park, K.~H.~Yoo, J.~Kim, H.~Y.~Ryu, and Q.~H.~Park, Appl. Phys. Lett.
\textbf{81}, 3239 (2002).
%15
\bibitem{plasmonic_metamaterials_JOPA05} F.~J.~Garcia-Vidal,
L.~Mart\'{i}n-Moreno, and J.~B.~Pendry, J. Opt. A: Pure Appl. Opt. \textbf{7}, S97 (2005).
%16
\bibitem{WoodHoles_PRB03} M.~Sarrazin, J.-P.~Vigneron, and
J.-M.~Vigoureux,  Phys. Rev. B \textbf{67},  085415 (2003).
%17
\bibitem{Popov_PRB00} E.~Popov, N.~Nevi\`{e}re, S.~Enoch, and
R.~Reinisch, Phys. Rev. B \textbf{62}, 16100 (2000).
%18
\bibitem{Book_Petit03} R.~Petit, M.~Neviere {\it Light propagation in periodic media. Differential Theory and Design},
(Marcel Dekker, New York, 2003).
%19
\bibitem{Zayats_PRL01} L.~Salomon, F.~Grillot, A.~V.~Zayats, and F.~de~Fornel,
Phys. Rev. Lett. \textbf{86}, 1110 (2001).
%20
\bibitem{Bloch_mode_OSA04} D.~G\'{e}rard, L.~Salomon, F.~de~Fornel, A.~V.~Zayats, Optics Express \textbf{12}, 3652 (2004).
%21
\bibitem{Grating_couplers_99} U.~Schr\"{o}ter and D.~Heitman,
Phys. Rev. B \textbf{60}, 4992 (1999).
%22
\bibitem{Nonzeroth_Barnes_APL_01} P.~T.~Worthing and
W.~L.~Barnes, Appl. Phys. Lett. \textbf{79},  3035 (2001).
%23
\bibitem{Arutsky_OptLett_00} I.~Arutsky, Y.~Zao, and
V.~Kochergin, Opt. Lett. \textbf{25}, 595 (2000).
%24
\bibitem{Popov2003_OptExpr} N.~Bonod, S.~Enoch, L.~Li, E.~Popov,
and M.~Nevi\`{e}re Opt. Expr. \textbf{11}, 428 (2003).
%25
\bibitem{LightTunelling_PhysRev} S.~A.~Darmanyan and
A.~V.~Zayats,  Phys. Rev. B \textbf{67}, 035424 (2003).
%26
\bibitem{Dykhne_PRB_03} A.~M.~Dykhne, M.~K.~Sarychev, and
V.~M.~Shalaev,  Phys. Rev. B \textbf{67}, 195402 (2003).
%27
\bibitem{AnalyticalTheory_PhysRev} S.~A.~Darmanyan, M.~Nevi\`{e}re, and
A.~V.~Zayats,  Phys. Rev. B \textbf{70}, 075103 (2004).
%28
\bibitem{Genchev_JETP04} Z.~D.~Genchev and D.~G.~Dosev,
J. Exp. Theor. Phys. \textbf{99}, 1129 (2004).
%29
\bibitem{Bliokh05} Yu.~P.~Bliokh, Phys. Rev. Lett. \textbf{95} , 165003 (2005).
%30
\bibitem{Franches_OptExpr05} A.~Benabbas, V.~Halt\'{e}, J.-Y.~Bigot,
Opt. Expr. \textbf{13}, 8730 (2005).
%31
\bibitem{Ours_JETPHL_04} A.~V.~Kats and A.~Yu.~Nikitin, JETP
Lett. \textbf{79}, 625 (2004).
%32
\bibitem{film_PhysRev} A.~V.~Kats and A.~Yu.~Nikitin,  Phys.~Rev.~B. \textbf{70}, 235412
(2004).
%33
\bibitem{KNN_PhysRev} A.~V.~Kats, M.~L.~Nesterov, and A.~Yu.~Nikitin,
Phys.~Rev.~B. \textbf{72}, 193405 (2005).
%34
\bibitem{Korea_03} S.~Park, G.~Lee, S.~H.~Song, C.~H.~Oh, and
P.~S.~Kim, Opt. Lett. \textbf{28}, 1870 (2003).
%35
\bibitem{LeePark_Nanoslits_PRB05} K.~G.~Lee, and
Q.-H. Park, Phys. Lett. \textbf{95}, 103902 (2005).
%36
\bibitem{GoldNanoparticlePlasmon_NANOL05} G.~Laurent, N.~F\'{e}lidj, S.~Lau~Truong, J.~Aubard, G. L\'{e}vi,
J.~R.~Krenn, A.~Hohenau, A.~Leitner, and F.~R.~Aussenegg, Nano Lett. \textbf{5}, 253 (2005).
%37
\bibitem{Nano_Chem_01} M.~D.~Malinsky, K.~L.~Kelly, G.~C.~Schatz, and
R.~P.~V.~Duyne, J. Am. Chem. Soc., \textbf{123}, 1471 (2001).
%38
\bibitem{NaomiHalas_MRS05} Y.~Xia and N.~J.~Halas, MRS Bulletin \textbf{30}, 338 (2005).
%39
\bibitem{hole_shape_PRL04} K.~J.~K.~Koerkamp, S.~Enoch, F.~B.~Segerink, N.~F.~Hulst, and L.~Kuipers,
Phys. Rev. Lett. \textbf{92}, 183901 (2004).
%40
\bibitem{LocalizedSPP_JOPA05} A.~Degiron and
 T.~Ebbesen, J. Opt. A: Pure Appl. Opt. \textbf{7}, S90 (2005).
%41
\bibitem{shape_localized_PRB05} K. L. van der Molen, K. J. Klein Koerkamp,
S. Enoch, F. B. Segerink, N. F. van Hulst, and L. Kuipers, Phys. Rev. B \textbf{72}, 045421 (2005).
%42
\bibitem{Gang_Sun_2006} Gang Sun and C. T. Chan, Phys. Rev. E \textbf{73}, 036613 (2006).
%43
\bibitem{Kelf_2006} T. A. Kelf, Y. Sugawara, R. M. Cole, and J. J. Baumberg, Phys. Rev. B \textbf{74}, 245415 (2006).
%44
\bibitem{local_SPP_in_voids} T. V. Teperik, V. V. Popov, F. Javier Garc\'{i}a de Abajo,
M. Abdelsalam, P. N. Bartlett, T. A. Kelf, Y. Sugawara, J. J. Baumberg, Opt. Expr. \textbf{14}, 1965 (2006).
%45
\bibitem{StrongPolariz_PRL04} R.~Gordon, A.~G.~Brolo, A.~McKinnon, A.~Rajora,
B.~Leathem, and K.~L.~Kavanagh,  Phys.~Rev.~Lett. \textbf{70}, 037401 (2004).
%46
\bibitem{PolarizControl_OptLett04} J.~Elliott, I.~I.~Smolyaninov, N.~I.~Zheludev and
A.~V.~Zayats,  Opt.~Lett. \textbf{29}, 1414 (2004).
%47
\bibitem{PolarizTomography_OptLett05} E.~Altewischer, C.~Genet, M.~P.~van~Exter, J.~P.~Woerdman,
 P.~F.~A.~Alkemade, A.~van~Zuuk and E.~W.~J.~M.~van~der~Drift, Opt. Lett. \textbf{30}, 90
 (2005).
%48
\bibitem{Opt.Depolariz_PRB05} C.~Genet, E.~Altewischer, M.~P.~van~Exter, and
J.~P.~Woerdman, Phys. Rev. B \textbf{71}, 033409 (2005).
%49
\bibitem{terahertz_Appl.Opt._04} F.~Miyamaru, and M.~Hangyo, Appl. Opt. \textbf{43},  1412 (2004).
%50
\bibitem{Plasmon_devices_05} S.~A.~Maier, Current Nanoscience,  \textbf{1}, 17 (2005).
%51
\bibitem{Jiri_Homola_1999} J. Homola, S. S. Yee, Gunter Gauglitz, Sensors and Actuators B \textbf{54}, 3 (1999).
%52
\bibitem{Jiri_Homola_2003} J. Homola, Anal. Bioanal. Chem.  \textbf{377}, 528 (2003).
%53
\bibitem{Paul_V_Lambeck_2006} P.~V.~Lambeck, Meas. Sci. Technol. \textbf{17}, R93 (2006).
%54
\bibitem{ELT_microwave_JOPA05} A.~G.~Schuchinsky, D.~E.~Zelenchuk, and A.~M.~Lerer, J. Opt. A: Pure Appl. Opt.,
\textbf{7}, S102 (2005).
%55
\bibitem{THz_SPP_diploma03} J.~Saxler, \textit{Surface Plasmon Polaritons at Terahertz
Frequencies on Metal and Semiuctor Surfaces} (Diploma Thesis, Aachen University, Aachen 2003).
%56
\bibitem{LRSPP_PRB91} F. Yang, J. R. Sambles and G. W. Bradberry, Phys. Rev. B \textbf{44}, 5855 (1991).
%57
\bibitem{Surface-Polariton-Like_PRB86} J. J. Burke, G. I. Stegeman and
T. Tamir, Phys. Rev. B \textbf{33}, 5186 (1986).
%58
\bibitem{SPIE} A.~V.~Kats and A.~Y.~Nikitin, Proc. SPIE \textbf{5221}, 218 (2003); \textbf{5477}, 381 (2004).
%59
\bibitem{Maradud_2D_gaps_PRB02} M.~Kretschmann, A.~A.~Maradudin, Phys. Rev. B \textbf{66}, 245408 (2002).
%60
\bibitem{ener.transfer_science04} P.~Andrew and W.~L.~Barnes,
Science \textbf{306}, 1002 (2004).
 %61
\bibitem{superlens_science05} N.~Fang, H.~Lee, C.~Sun, and X.~Zhang, Science \textbf{308}, 534
(2005).
%62
\bibitem{Book_Zolotarev84} V.~M.~Zolotarev, V.~N.~Morozov, and E.~V.~Smirnova
 {\it Optical constants of natural media}, (Chemistry, Leningrad, 1984).
%63
\bibitem{Landau} L.~D.~Landau and E.~M.~Lifshits, {\it Classical Mechanics} (Pergamon, Oxford, 1977).


\end{thebibliography}
\end{document}